\documentclass[onecolumn,amsthm,reviewcopy]{arxiv-preprint}
\usepackage{amsmath,amsfonts,amssymb}
\usepackage{mathrsfs}
\usepackage{graphicx}
\usepackage{xcolor,booktabs,array,url}
\usepackage[authoryear,round]{natbib}
\usepackage{tikz}
\usetikzlibrary{arrows.meta,positioning}

\ifdim\paperheight=0pt\relax
\fi
\usepackage[
    colorlinks=true,
    citecolor=blue,
    linkcolor=black,
    urlcolor=blue,
    setpagesize=false
]{hyperref}

\AddToHook{env/frontmatter/after}{\endNoHyper}

\newtheorem{theorem}{Theorem}
\newtheorem{proposition}[theorem]{Proposition}

\begin{document}

\begin{frontmatter}
\date{16 November 2026}
\title{Robust Data-Driven Output-Feedback Reference Governor with Quantified Model Uncertainty}
\author{Tam W. Nguyen}\ead{nguyen.tamwilly.3e@kyoto-u.ac.jp}
\address{Department of Electrical Engineering, Kyoto University, Kyoto 615-8510, Japan}
\begin{abstract}
This paper constructs a data-driven reference governor (DDRG) from
input--output data and supplied deterministic disturbance bounds.
A data-consistent model set quantifies the uncertainty remaining in
each prediction direction.
With exact output measurements, the DDRG propagates output-history
uncertainty while retaining exact applied-reference histories and
converts the resulting tubes into constraint margins.
Under a common contraction condition in a selected history norm,
explicit terminal feasibility inequalities certify infinite-horizon
continuation under a held reference and yield a closed invariant
admissible set.
A scalar update selects the largest certified step toward the requested
reference, with recursive feasibility and constraint satisfaction from
certified initialization.
A separate bounded-measurement-noise extension derives the
errors-in-variables model set and the additional physical-output
margin through a measured-history certificate.
On the coupled benchmark, longer nested records reduce one-step
prediction disagreement at the probe regressor, and propagation
accounts for most of the gap between the tube and true-model
envelopes at the reported constraint row and prediction step.
\end{abstract}

\end{frontmatter}

\section{Introduction}
A reference governor manages constraints by modifying the reference
supplied to an already stabilized closed-loop system. Each update must
account for the immediate response and safe continuation. Maximal
output admissible sets characterize initial conditions whose subsequent
outputs satisfy the constraints. For unforced linear systems,
\citet{gilbert1991maximal} give finite representations
or approximations under suitable conditions.
\citet{kolmanovsky1995maximal} characterize maximal output admissible sets
for discrete-time linear systems with bounded disturbances
and develop methods for computing maximal disturbance-invariant
sets~\citep{kolmanovsky1998theory}.
A reference governor tests admissibility jointly for the initial
condition and a candidate reference held constant during prediction.
\citet{gilbert1999fast} develop this construction
for bounded disturbances and establish constraint satisfaction from
admissible initialization. The option to hold a previously certified
reference gives a feasible decision at the next sample.
In predictive command management, \citet{bemporad1997command} use
virtual command sequences closed under time shifts to preserve
feasibility and enforce constraints from admissible initialization.
For linear periodic systems, \citet{ossareh2020periodic}
develops periodically invariant maximal output admissible sets and
reference governors.

With incomplete state information, the admissibility test must account
for uncertainty in the initial condition.
\citet*{angeli2001feasible} study command governors with set-membership
state estimation and show how approximating the state uncertainty set
affects repeated feasibility. In robust output-feedback model predictive
control (MPC), \citet{mayne2006robust} use invariant sets to
bound estimation and control errors. They tighten nominal constraints to
protect the physical state and input. For nonlinear systems with
state- and input-dependent uncertainty, \citet{koehler2021computational}
construct prediction-dependent tubes and establish recursive feasibility
by showing that the candidate tube lies inside the previous optimal tube.
These constructions link uncertainty containment and constraint tightening
to feasibility at successive samples.

Reference governors for uncertain systems also use estimation,
learning, and uncertainty compensation. \citet{dongare2023reference}
combine ultra-local modeling with an observer for unknown dynamics.
They use the observer's ultimate error bound to construct robust
admissible sets. \citet*{chakrabarty2023parameter}
use offline simulations and online parameter confidence regions to
learn parameter-robust constraint-admissible sets. Their probabilistic
guarantee of closed-loop constraint satisfaction depends on their
estimation and classification assumptions. \citet*{zhao2024integrated}
combine an $\mathcal L_1$ adaptive controller with a reference governor
for known nominal dynamics. Uniform bounds on the differences between
actual and nominal state and input trajectories determine the constraint
tightening. \citet{castroviejofernandez2024less}
compensate in the nominal control law for bounded disturbances
estimated from outputs. Their time-dependent admissible sets use
estimation error bounds. These designs base admissibility on estimation
error bounds, parameter confidence regions, or certified deviations
from nominal dynamics. \citet{portelladelgado2026extremum}
numerically study a related adaptive architecture combining online
input--output identification, constrained predictive control, and
extremum-seeking command generation.

Data-driven formulations also supply predictions for admissibility
tests. \citet{masti2020neural} jointly design a data-driven low-level
controller and a receding-horizon neural reference governor for
nonlinear systems. For behavioral prediction, the fundamental lemma of
\citet{willems2005persistency} shows that windows of a trajectory with
sufficiently persistently exciting input span the finite-length
behavior of a controllable linear time-invariant system.
\citet{ossareh2023maximal} uses behavioral and subspace
predictors to construct a data-driven counterpart of a finitely
determined inner approximation of the maximal admissible set from
noise-free input--output records, and establishes its relation to the
model-based formulation under the stated conditions. The reference
governor operates directly on recent input--output histories without
reconstructing a state-space realization.
For nonlinear closed-loop systems, \citet*{shi2025koopman}
combine Koopman lifting, behavioral prediction, and a data-driven
admissible set with robust reference optimization over prescribed
uncertainty in the predictor data.

Robust behavioral MPC addresses uncertainty in recorded trajectories
and online measurements. \citet{berberich2021ddmpc}
establish practical exponential stability for multistep input--output
MPC with bounded output-measurement noise, regularization, and
terminal equality constraints, under suitable excitation, noise, and
design conditions. Their constraint-satisfaction
study~\citep{berberich2020robustconstraints} uses output tightening to
establish recursive feasibility and closed-loop output-constraint
satisfaction under the stated conditions. \citet{kloeppelt2025constrainttightening}
derive constraint tightening for process disturbances, with
prestabilization and an output-feedback formulation that uses a
finite-lag input--output representation. \citet{kerz2023tubestochastic}
combine data-driven nominal and error dynamics with stochastic tubes,
robust treatment of bounded online measurement errors, terminal
ingredients, and a robust first-step constraint. Their guarantees use
exact offline input, state, and disturbance records. The state
constraints are satisfied in the chance-constrained sense, and
recursive feasibility is robust.

A model-consistency approach retains the set of models compatible with
the data and supplied uncertainty bounds. \citet{tanaskovic2014adaptive}
combine recursive set-membership identification with predictive
control for open-loop stable multiple-input multiple-output systems.
From feasible initialization, their controller enforces output
constraints for all retained models and remains feasible as the model
set contracts. \citet*{xie2026data}
characterize compatible system matrices from input--state data
affected by bounded process noise. They use semidefinite optimization
to obtain a min--max state-feedback controller with robust closed-loop
guarantees. In both cases, the model set makes finite-data uncertainty
explicit in the constrained decision.
A complementary line combines model learning with invariant-set
construction. \citet{mulagaleti2022invariant} jointly select an uncertain
prediction model, a feedback gain, and tube and terminal invariant sets
for robust MPC from state--input data. \citet{mulagaleti2026lpv}
integrate quasi-LPV model learning from input--output data with
observer-based uncertainty descriptions and robust control invariant-set
constraints.
\citet{bernstein2022facing} discusses how modeling assumptions
shape uncertainty descriptions and how feedback-control experiments
can probe physical variations.

We use model consistency to examine how retained uncertainty affects a
reference-governor certificate. At a fixed history and held reference,
a propagated tube can include points outside the trajectories of all
compatible models with fixed coefficients. A tube-based test can then
fail even if every compatible fixed-model trajectory satisfies the
constraint. We examine how retained model uncertainty and propagation
overestimation consume constraint margin, and when the resulting
bounds certify a reference change. Reference authority is the largest
certified fraction of the segment from the previous applied reference
to the current request.

We construct a data-driven reference governor (DDRG) for a fixed
stabilizing controller. Its closed loop admits a constant-coefficient,
finite-lag input--output recurrence with a known sufficient history
length. The same unknown coefficients govern data collection and
operation. The assumptions include stability of the zero-reference,
zero-disturbance recurrence, unit steady-state tracking of the
designated outputs, and a deterministic process-disturbance bound.
The data-consistent model set retains the coefficient matrices that fit
the recorded transitions within the disturbance bound and obey any
supplied coefficient restrictions. The exact-measurement construction
assumes a compact model set. The DDRG keeps this set, a feasible nominal
model, and the selected history norm fixed during online operation.
The tube propagates uncertainty in the output history. The stored applied
references update exactly. A separate extension treats bounded
output-measurement errors in the offline record and online history.

The DDRG combines set-membership uncertainty bounds, robust tube
propagation, and admissibility-based updates. Its contributions
concern the certificate, its conservatism, and the extension to
noisy measurements.

\begin{enumerate}
\item \textbf{Data-consistent held-reference certification.}
We bound one-step prediction disagreement in directions supported by
the recorded regressors and state conditions for compactness of the
model set. With a fixed compact model set, the output-history tube
accounts for nominal propagation, local model mismatch, and process
disturbances. We prove trajectory containment and show that the
shifted preceding tube contains the entire restarted tube. Under the
additional contraction condition and for the specified horizons,
the stage, terminal-radius-floor, and terminal-entry inequalities
certify indefinite continuation under a held reference. They
eliminate the terminal radius exactly over the specified family,
including zero radius, and define a closed admissible set that is
invariant under the held reference. The admissible sets are nested
in the prediction horizon and are convex when the applied-reference
set is convex.

\item \textbf{Constraint-margin accounting and matched reference authority.}
For each constraint row and prediction step, we compare the
true-model envelope, the envelope over compatible models whose
coefficients remain fixed throughout each rollout, and the propagated
tube envelope. All use the same initial history, held reference, and
disturbance set. Their ordering separates the gap from retained model
uncertainty from the additional gap due to tube geometry and
propagation bounds. The true-model envelope is an evaluation
diagnostic available when the plant is known. The online DDRG uses
the tube envelope. We compare output-history tubes with full-history
error balls under fixed data, model, disturbance, constraint, and
horizon choices and the stated norm compatibility condition. The
resulting admissible-set inclusion orders the largest certified reference
fractions pointwise, with equality allowed.

\item \textbf{Physical-output certification with bounded measurement noise.}
In the errors-in-variables consistency description, each offline
output sample uses the same sensor-error witness wherever it occurs.
The offline error bound enters the consistency test. The online error
bound and the process-disturbance bound give a forcing bound for
measured-history prediction over the retained models. This bound
covers every admissible completion of the next sensor error. This
coverage converts surrogate admissibility into physical-output
constraint satisfaction and preserves the restart and held-reference
continuation arguments. The guarantee requires a fixed certified
compact model set, a verified consistent nominal, a fixed history norm,
the corresponding contraction condition, an admissible measured-history
hold, and a separate physical-safety certificate at initialization.
\end{enumerate}

Containment and restart hold for any nonnegative propagation gain
$q$. This gain is the sum of the nominal output-history gain and the
uniform model-deviation allowance in the selected norm. The terminal
construction requires $q<1$, which can fail even when the true plant
is stable. From certified initialization, the largest certified
segment update is attained, and a feasible hold is available at
subsequent samples. The DDRG guarantees recursive feasibility and
satisfaction of the physical-output and applied-reference constraints
at sampling instants. Eventual attainment of an arbitrary request
lies outside this guarantee.

A coupled lag-two benchmark examines how retained model uncertainty
affects margins and reference updates. Longer nested records reduce
one-step prediction disagreement at the common nominal model and
query. At the reported constraint row and prediction step,
propagation accounts for most of the gap between the tube and
true-model envelopes. Matched comparisons evaluate the effects of
output-history propagation and terminal-radius elimination on
certified reference fractions. Exact-measurement and bounded-noise
runs illustrate the corresponding initialization and safety conditions.
The model family shares an equilibrium and steady-state gain, so the
comparisons concern transient uncertainty and the specified
certificates.

Section~\ref{sec:problem} states the plant, information, and constraint
assumptions. Section~3 develops the data-consistent uncertainty bounds,
prediction tubes, margin analysis, terminal certificate, matched
comparisons, and reference update. Section~\ref{sec:measurement-noise}
treats bounded output-measurement noise. Section~\ref{num:section}
presents the numerical investigation, with parameters and
certification details in Appendix~\ref{num:appendix}.

\providecommand{\coloneqq}{\mathrel{:=}}

\section{Problem setup}
\label{sec:problem}

\subsection*{Notation}
The operator \(\operatorname{col}\) stacks its arguments vertically.
For a signal indexed by time, the notation \(x_{a:b}\) denotes its
samples from \(a\) through \(b\), inclusive.
For a finite real vector \(x\), we use the norms
\[
    \|x\|_1=\sum_i|x_i|,\qquad
    \|x\|_2=\Bigl(\sum_i x_i^2\Bigr)^{1/2},\qquad
    \|x\|_\infty=\max_i|x_i|.
\]
For a real matrix \(A=(A_{ij})\), the entrywise maximum norm is
\(\|A\|_{\max}=\max_{i,j}|A_{ij}|\) and the Frobenius norm is
\(\|A\|_{\mathrm{F}}=(\sum_{i,j} A_{ij}^2)^{1/2}\).
The matrix norms \(\|A\|_2\) and \(\|A\|_\infty\) are induced by the
corresponding vector norms. In particular, the induced infinity
norm is the largest absolute row sum.
Vector inequalities are componentwise.

\subsection{System and prior knowledge}
The fixed stabilizing controller receives the applied reference.
Assume that the resulting discrete-time closed loop admits the
input--output representation, with known sufficient history length
\(\ell\),
\begin{equation}
    y_{k+1}
    =
    \sum_{j=0}^{\ell-1} F_j y_{k-j}
    +
    \sum_{j=0}^{\ell-1} G_j v_{k-j}
    +w_k,
    \label{eq:system}
\end{equation}
where $y_k\in\mathbb R^p$ is the physical output,
$v_k\in\mathbb R^m$ is the applied reference, and
$w_k\in\mathbb R^p$ is the disturbance over the transition from
\(k\) to \(k+1\). The output and reference dimensions \(p\) and \(m\),
and the history length \(\ell\), are known positive integers.
The matrices $F_j\in\mathbb R^{p\times p}$ and
$G_j\in\mathbb R^{p\times m}$ are unknown and constant.
The same coefficients govern data collection and online operation.
Assume that with the applied reference and disturbance identically zero
in \eqref{eq:system}, the output tends to zero for every initial output
history.

Let $r_k\in\mathbb R^m$ denote the requested reference and let
$y_k^{\mathrm{tr}}=Ty_k$ be the tracked output,
where $T\in\mathbb R^{m\times p}$ is known.
For a constant applied reference $\bar v$ and zero disturbance, the
steady-state physical output \(\bar y\) is given by
\begin{equation}
    \bar y=G_{\mathrm{dc}}\bar v,
    \qquad
    G_{\mathrm{dc}}\coloneqq
    \left(I_p-\sum_{j=0}^{\ell-1}F_j\right)^{-1}
    \sum_{j=0}^{\ell-1}G_j,
    \label{eq:steady-state}
\end{equation}
where \(G_{\mathrm{dc}}\in\mathbb R^{p\times m}\) is the steady-state gain.
Note that a nonzero vector in the kernel of \(I_p-\sum_j F_j\) would
give a constant nonzero output with zero applied reference and
disturbance, contradicting the assumed convergence in \eqref{eq:system}.
The matrix therefore has an inverse.
Assume unit steady-state tracking,
\begin{equation}
    TG_{\mathrm{dc}}=I_m.
    \label{eq:tracking}
\end{equation}
It follows that $y_k^{\mathrm{tr}}\to\bar v$ when the applied
reference is held at $\bar v$ and $w_k=0$.

The supplied disturbance bound determines the set of admissible disturbances
\begin{equation}
    w_k\in\mathcal W
    \coloneqq
    \{w\in\mathbb R^p:\|w\|_\infty\le\bar w\},
    \qquad \bar w\ge0,
    \label{eq:disturbance}
\end{equation}
during both data collection and online operation. The analysis assumes
no disturbance distribution.

\subsection{Available data and measurements}
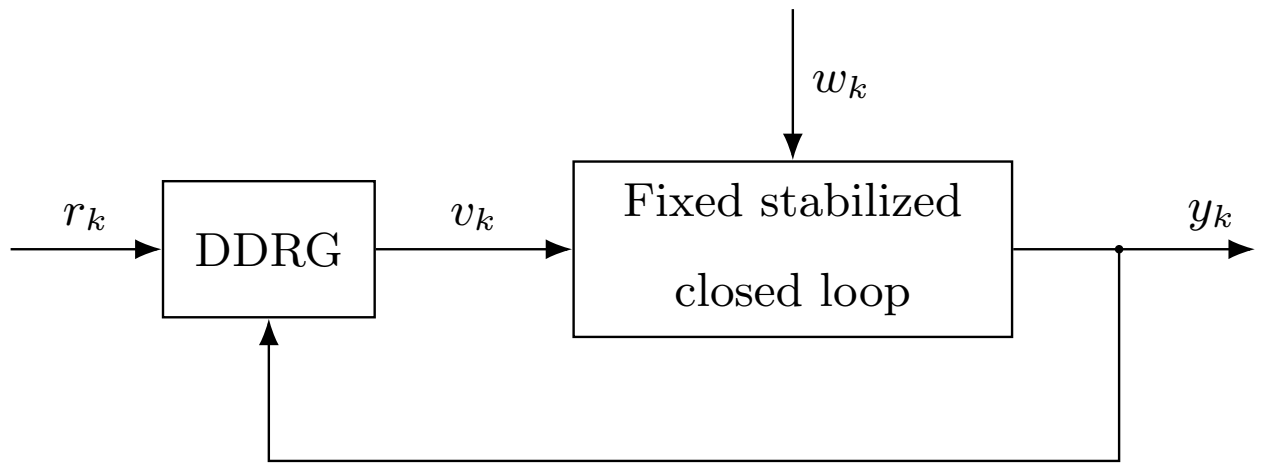
\begin{figure}[htbp]
  \centering
  \begin{tikzpicture}[
    font=\small,
    >=Latex,
    line width=0.45pt,
    ddrgblock/.style={draw,align=center,minimum height=9mm,inner sep=4pt}
  ]
    \node[ddrgblock,minimum width=14mm] (rg) {DDRG};
    \node[ddrgblock,minimum width=29mm,right=13mm of rg] (cl)
      {Fixed stabilized\\closed loop};

    \draw[->] ([xshift=-10mm]rg.west)
      -- node[above] {$r_k$} (rg.west);
    \draw[->] (rg.east)
      -- node[above] {$v_k$} (cl.west);
    \draw[->] (cl.east)
      -- ++(16mm,0) node[above left] {$y_k$};

    \coordinate (out) at ([xshift=7mm]cl.east);
    \fill (out) circle (0.8pt);
    \draw[->] (out) -- ++(0,-14mm) -| (rg.south);
    \draw[->] ([yshift=10mm]cl.north)
      -- node[right] {$w_k$} (cl.north);
  \end{tikzpicture}
  \caption{Reference-governor architecture with exact output measurements.
  The DDRG selects $v_k$ from the request $r_k$ using the output and
  stored applied-reference histories. The combined block contains the
  plant and fixed stabilizing controller. The disturbance $w_k$ acts
  over the transition from $k$ to $k+1$.}
  \label{fig:ddrg-architecture}
\end{figure}

The offline record pairs applied references with output samples from
the same closed loop in the dataset
\begin{equation}
    \mathcal D\coloneqq (v_{\mathrm{d},0:N-1},y_{\mathrm{d},0:N}),
    \qquad N\in\mathbb N,\quad N\ge\ell,
    \label{eq:dataset}
\end{equation}
where \(N\) is the number of recorded transitions and
the upright subscript \(\mathrm{d}\) marks offline data.
The baseline data-driven reference governor (DDRG) uses exact
applied-reference records and exact output measurements to isolate how
model uncertainty consumes constraint margin.
At online initialization $k=0$, the required output history
$y_{-\ell+1:0}$ and the past applied-reference history are available.
The previous reference $v_{-1}$ is also available, including when
$\ell=1$ and the dynamic past-reference block is empty.
Figure~\ref{fig:ddrg-architecture} shows the exact-measurement architecture.

Section~\ref{sec:measurement-noise} treats sensor noise separately.
It retains the same reference governor architecture and derives
conditions on identification, prediction, and physical-output safety
for distinct offline and online measurement-error bounds.

\subsection{Constraints and objective}
At each sampling instant, the physical output and the applied reference
must satisfy
\begin{equation}
    Hy_k\le h,\qquad v_k\in\mathcal V,
    \label{eq:constraints}
\end{equation}
where $H\in\mathbb R^{n_c\times p}$,
$h\in\mathbb R^{n_c}$, and the compact set
$\mathcal V\subset\mathbb R^m$ are known.
The DDRG acts on the reference supplied to the fixed controller.
Starting from a certified initial history, it must enforce
\eqref{eq:constraints} for every disturbance sequence in
\eqref{eq:disturbance}. At each update it selects the largest certified
step along the segment from the previous applied reference to the
current request.

The certificate bounds the constraint margin consumed by model
uncertainty retained by the recorded data and checks whether the
remaining margin suffices for a reference change. The five stages are
\[
\begin{gathered}
\text{Data consistency}\\
\downarrow\\
\text{Prediction uncertainty}\\
\downarrow\\
\text{Constraint margins}\\
\downarrow\\
\text{Invariant admissibility}\\
\downarrow\\
\text{Safe reference updates}
\end{gathered}
\]
The terminal certificate extends finite predictions to indefinite
continuation of the entire prediction tube. This continuation preserves
a feasible hold action for the next update.

\section{Data-Driven Uncertainty Quantification and Constraint Certification}

\subsection{Data-consistent prediction uncertainty}
\label{sec:data}

Write \eqref{eq:system} in regression form to separate the fixed unknown
coefficients from the output and applied-reference coordinates in
\begin{equation}
    y_{k+1}=\Theta\phi_k+w_k,
    \label{eq:regression}
\end{equation}
where
\begin{align}
    \Theta
    &\coloneqq
    \begin{bmatrix}
        F_0 & \cdots & F_{\ell-1} &
        G_0 & \cdots & G_{\ell-1}
    \end{bmatrix},\\
    \phi_k
    &\coloneqq
    \operatorname{col}
    \left(
        y_k,\ldots,y_{k-\ell+1},
        v_k,\ldots,v_{k-\ell+1}
    \right)
    \in\mathbb R^d,
\end{align}
with $d=\ell(p+m)$ and unknown
$\Theta\in\mathbb R^{p\times d}$.

Extract the $N_{\mathrm{d}}=N-\ell+1$ valid transitions, indexed by
$k=\ell-1,\ldots,N-1$, from
the offline dataset $\mathcal D$ and define
\begin{equation}
    Y\coloneqq
    \begin{bmatrix}
        y_{\mathrm{d},\ell} & \cdots & y_{\mathrm{d},N}
    \end{bmatrix},
    \quad
    \Phi\coloneqq
    \begin{bmatrix}
        \phi_{\mathrm{d},\ell-1} & \cdots & \phi_{\mathrm{d},N-1}
    \end{bmatrix},
\end{equation}
where $Y\in\mathbb R^{p\times N_{\mathrm{d}}}$,
$\Phi\in\mathbb R^{d\times N_{\mathrm{d}}}$, and each
$\phi_{\mathrm{d},i}$ is formed from the recorded output and
applied-reference history at time $i$.
Let $W=[\,w_{\mathrm{d},\ell-1}\ \cdots\ w_{\mathrm{d},N-1}\,]
\in\mathbb R^{p\times N_{\mathrm{d}}}$ collect the corresponding offline
disturbances. The recorded transitions satisfy
\begin{equation}
    Y=\Theta\Phi+W,
    \qquad
    \|W\|_{\max}\leq\bar w.
    \label{eq:data-equation}
\end{equation}

The prior information is a known closed set
$\mathcal K\subseteq\mathbb R^{p\times d}$ containing $\Theta$.
Use $\mathcal K=\mathbb R^{p\times d}$ when no coefficient
restrictions are supplied.

\begin{proposition}[Data-consistent prediction uncertainty]
\label{prop:data}
\leavevmode
\begin{enumerate}
\item \textit{Containment and prediction.}
The data-consistent model set collects every candidate matrix that
satisfies the prior restrictions and fits all recorded transitions
within the supplied disturbance bound. Define it by
\begin{equation}
    \mathcal P
    \coloneqq
    \left\{
        M\in\mathcal K:
        \|Y-M\Phi\|_{\max}\leq\bar w
    \right\}.
    \label{eq:model-set}
\end{equation}
Then $\Theta\in\mathcal P$.

Choose any feasible member $\widehat\Theta\in\mathcal P$ as the nominal
model. To quantify the disagreement between the nominal one-step
prediction and those of the retained models at a fixed regressor
\(\phi\in\mathbb R^d\), define
\begin{equation}
    \epsilon(\phi)
    \coloneqq
    \sup_{M\in\mathcal P}
    \|(M-\widehat\Theta)\phi\|_\infty.
    \label{eq:model-uncertainty}
\end{equation}
The infinity norm takes the largest absolute output-component difference
at the fixed regressor.
When $\mathcal P$ is unbounded, the supremum can equal $+\infty$, indicating that the
available information gives no finite bound in that direction.
Adding the disturbance bound gives, for every admissible transition,
\begin{equation}
    \|y_{k+1}-\widehat\Theta\phi_k\|_\infty
    \leq\epsilon(\phi_k)+\bar w.
    \label{eq:prediction-bound}
\end{equation}

\item \textit{Directional bound.}
For $\phi\in\operatorname{range}(\Phi)$, the column space of \(\Phi\),
represent \(\phi\) as a linear combination of recorded regressors.
The vector \(c\in\mathbb R^{N_{\mathrm d}}\) contains one real
linear-combination weight per recorded regressor, with \(\Phi c=\phi\)
as its only constraint. To choose the representation that minimizes
the residual bound in the proof, define
\begin{equation}
    \chi_{\mathrm{d}}(\phi)
    \coloneqq
    \min_{\Phi c=\phi}\|c\|_1.
    \label{eq:data-coverage}
\end{equation}
Then
\begin{equation}
    \epsilon(\phi)\leq2\bar w\,\chi_{\mathrm{d}}(\phi).
    \label{eq:directional-bound}
\end{equation}

\item \textit{Full row rank.}
If $\Phi$ has full row rank, $\mathcal P$ is compact,
the supremum in \eqref{eq:model-uncertainty} is attained,
and
\begin{equation}
    \epsilon(\phi)
    \leq
    \frac{2\bar w\sqrt{N_{\mathrm{d}}}}{\sigma_{\min}(\Phi)}
    \|\phi\|_2,
    \qquad \forall\phi\in\mathbb R^d,
    \label{eq:excitation-bound}
\end{equation}
where \(\sigma_{\min}(\Phi)\) is the smallest singular value of \(\Phi\).
\end{enumerate}

\end{proposition}

\begin{proof}
    \textit{Model containment and one-step prediction.}
    From \eqref{eq:data-equation},
    \[
        \|Y-\Theta\Phi\|_{\max}
        =\|W\|_{\max}\leq\bar w.
    \]
    Since $\Theta\in\mathcal K$, it follows that
    $\Theta\in\mathcal P$.
    For any admissible transition,
    \begin{align*}
        \|y_{k+1}-\widehat\Theta\phi_k\|_\infty
        &=
        \|(\Theta-\widehat\Theta)\phi_k+w_k\|_\infty\\
        &\leq
        \|(\Theta-\widehat\Theta)\phi_k\|_\infty
        +\|w_k\|_\infty\\
        &\leq\epsilon(\phi_k)+\bar w,
    \end{align*}
    which establishes \eqref{eq:prediction-bound}.

    \textit{Directional uncertainty.}
    For $M\in\mathcal P$, define the residuals
    \[
        R_M\coloneqq Y-M\Phi,
        \qquad
        \widehat R\coloneqq Y-\widehat\Theta\Phi.
    \]
    Both residuals satisfy
    $\|R_M\|_{\max},\|\widehat R\|_{\max}\leq\bar w$,
    and
    \[
        (M-\widehat\Theta)\Phi=\widehat R-R_M.
    \]
    For any residual matrix \(R\in\mathbb R^{p\times N_{\mathrm d}}\)
    with \(\|R\|_{\max}\leq\bar w\), the absolute value of each row
    applied to \(c\) is bounded by \(\bar w\sum_i|c_i|\), giving
    \(\|Rc\|_\infty\leq\bar w\|c\|_1\). Applying this bound to both
    residuals, for any $c$ satisfying $\Phi c=\phi$ we obtain
    \begin{align*}
        \|(M-\widehat\Theta)\phi\|_\infty
        &=
        \|(\widehat R-R_M)c\|_\infty\\
        &\leq
        \|\widehat R c\|_\infty+\|R_Mc\|_\infty\\
        &\leq2\bar w\|c\|_1.
    \end{align*}
    Taking the supremum over $M\in\mathcal P$ and minimizing
    over $\Phi c=\phi$ yields \eqref{eq:directional-bound}.
    The minimum is attained because, for any feasible $c_0$,
    the search can be restricted to the nonempty compact set
    \[
        \{c:\Phi c=\phi,\ \|c\|_1\leq\|c_0\|_1\}.
    \]

    \textit{Full row rank.}
    Suppose $\operatorname{rank}(\Phi)=d$.
    Let \(\Phi^\dagger\) denote the Moore--Penrose inverse.
    With the matrix norm induced by the Euclidean vector norm,
    \[
        \Phi\Phi^\dagger=I_d,
        \qquad
        \|\Phi^\dagger\|_2
        =\frac{1}{\sigma_{\min}(\Phi)}.
    \]
    For every $M\in\mathcal P$,
    \[
        M-\widehat\Theta
        =
        (\widehat R-R_M)\Phi^\dagger.
    \]
    Converting the entrywise residual bounds to the Frobenius norm gives
    \[
        \|M-\widehat\Theta\|_{\mathrm{F}}
        \leq
        \frac{2\bar w\sqrt{pN_{\mathrm{d}}}}{\sigma_{\min}(\Phi)},
    \]
    so $\mathcal P$ is bounded.
    Since $\mathcal K$ is closed and the data inequalities are
    continuous, $\mathcal P$ is also closed, hence compact.
    Continuity of
    $M\mapsto\|(M-\widehat\Theta)\phi\|_\infty$
    then ensures that the supremum defining $\epsilon(\phi)$
    is attained.

    Finally, $c=\Phi^\dagger\phi$ satisfies $\Phi c=\phi$.
    Therefore,
    \begin{align*}
        \chi_{\mathrm{d}}(\phi)
        &\leq\|\Phi^\dagger\phi\|_1\\
        &\leq\sqrt{N_{\mathrm{d}}}\|\Phi^\dagger\phi\|_2\\
        &\leq
        \frac{\sqrt{N_{\mathrm{d}}}}{\sigma_{\min}(\Phi)}
        \|\phi\|_2.
    \end{align*}
    Combining this with \eqref{eq:directional-bound}
    proves \eqref{eq:excitation-bound}.
\end{proof}

The system equation \eqref{eq:system}, disturbance bound
\eqref{eq:disturbance}, and prior restriction \(\Theta\in\mathcal K\)
ensure that the model set is nonempty because it contains the true
model. A candidate from a least-squares fit is eligible as a nominal
once its membership in \(\mathcal P\) is verified.

The directional quantity \(\chi_{\mathrm d}(\phi)\) uses information
from the recorded regressors at one fixed \(\phi\). Finite uncertainty
bounds in specific directions can coexist with an unbounded model
set. The prediction tubes use maxima over the entire model set and
require compactness. Full row rank is sufficient for compactness.
With rank-deficient data, a justified coefficient parameterization
or bounds in \(\mathcal K\) can still give boundedness, which together
with closedness gives compactness.
For a polyhedral model set, finite minima and maxima of every
coefficient, obtained by linear programming, give a boundedness
certificate. Rank deficiency can reflect insufficient regressor
variation or arise structurally from an excessive lag, so the
parameterization and available coefficient information must be
considered together with the input data.

\subsection{Certified prediction tubes}

The output history depends on uncertain coefficients and disturbances.
The past applied references are stored and update exactly when a new
reference is applied. Let \(L\coloneqq\ell-1\),
\(n_{\mathrm{o}}\coloneqq\ell p\), and \(n\coloneqq n_{\mathrm{o}}+Lm\).
The output history, applied-reference history, and stacked history are
\[
\begin{aligned}
    z_k&\coloneqq\operatorname{col}(y_k,\ldots,y_{k-L})
        \in\mathbb R^{n_{\mathrm{o}}},\\
    u_k&\coloneqq\operatorname{col}(v_{k-1},\ldots,v_{k-L})
        \in\mathbb R^{Lm},\\
    \xi_k&\coloneqq\operatorname{col}(z_k,u_k)\in\mathbb R^n.
\end{aligned}
\]
The vector \(u_k\) stores references applied to the fixed controller
and is empty when \(\ell=1\). For \(\xi=\operatorname{col}(z,u)\) and
\(v\in\mathbb R^m\), the regressor
\[
    \phi(\xi,v)\coloneqq\operatorname{col}(z,v,u)\in\mathbb R^d
\]
stacks outputs, the current applied reference, and past applied
references in the coefficient order of \eqref{eq:regression}.

For a candidate \(M\in\mathcal P\), the history dynamics are
\begin{equation}
\begin{aligned}
    z^+&=\mathsf A_Mz+\mathsf D_Mu+\mathsf B_Mv+E_{\mathrm{o}}w,\\
    u^+&=Su+Jv,\qquad y=C_{\mathrm{o}}z,
\end{aligned}
\label{eq:candidate-history-dynamics}
\end{equation}
where \(+\) denotes the next sample. The maps \(\mathsf A_M\),
\(\mathsf D_M\), and \(\mathsf B_M\) act on the output history, past
applied references, and current applied reference, respectively.
The map \(E_{\mathrm{o}}\) injects the disturbance into the first
block, and \(C_{\mathrm{o}}\) reads the current output from that
block. The maps \(S,J\) shift and insert applied references.

To specify the maps, partition the candidate coefficients as
\[
    M=[\,F_0(M)\ \cdots\ F_L(M)\ G_0(M)\ \cdots\ G_L(M)\,].
\]
For \(L\geq1\), write \(z=\operatorname{col}(z^{(0)},\ldots,z^{(L)})\) and
\(u=\operatorname{col}(u^{(1)},\ldots,u^{(L)})\), with
\(z^{(r)}\in\mathbb R^p\) and \(u^{(r)}\in\mathbb R^m\).
The operators form the new output in the first block and retain the
\(L\) most recent output samples in the remaining blocks through
\[
\begin{aligned}
    \mathsf A_Mz
    &\coloneqq\operatorname{col}\Bigl(
         \sum_{r=0}^{L}F_r(M)z^{(r)},z^{(0)},\ldots,z^{(L-1)}\Bigr),\\
    \mathsf D_Mu
    &\coloneqq\operatorname{col}\Bigl(
         \sum_{r=1}^{L}G_r(M)u^{(r)},0_{Lp}\Bigr),\\
    \mathsf B_Mv
    &\coloneqq\operatorname{col}\bigl(G_0(M)v,0_{Lp}\bigr),\qquad
    E_{\mathrm{o}}w\coloneqq\operatorname{col}(w,0_{Lp}),\qquad
    C_{\mathrm{o}}z\coloneqq z^{(0)}.
\end{aligned}
\]
The dimensions of \(\mathsf A_M,\mathsf D_M,\mathsf B_M\) are
\(n_{\mathrm{o}}\times n_{\mathrm{o}}\), \(n_{\mathrm{o}}\times Lm\),
and \(n_{\mathrm{o}}\times m\), respectively.
The maps \(E_{\mathrm{o}}\) and \(C_{\mathrm{o}}\) have dimensions
\(n_{\mathrm{o}}\times p\) and \(p\times n_{\mathrm{o}}\).
For \(L\geq1\), the \(m\times m\) blocks of
\(S\in\mathbb R^{Lm\times Lm}\) are \(S_{ab}=I_m\) when \(a=b+1\)
and zero otherwise, and
\(J\coloneqq\operatorname{col}(I_m,0_{(L-1)m\times m})
\in\mathbb R^{Lm\times m}\). Thus the reference-history update is
\[
    Su+Jv=\operatorname{col}(v,u^{(1)},\ldots,u^{(L-1)}),
    \qquad S^L=0.
\]
When \(L=0\), omit the lower output-shift rows and set
\(\mathsf A_M=F_0(M)\), \(\mathsf B_M=G_0(M)\), and
\(E_{\mathrm{o}}=C_{\mathrm{o}}=I_p\). The dimensions of
\(\mathsf D_M,S,J\) are then \(p\times0\), \(0\times0\), and
\(0\times m\), respectively.

The same dynamics can act on the full history \(\xi\). To write
the successor \(\xi^+=\mathsf{F}_M\xi+\mathsf{G}_Mv+Ew\) and
output \(y=C\xi\), define
\[
\begin{aligned}
    \mathsf{F}_M&\coloneqq
    \begin{bmatrix}\mathsf A_M&\mathsf D_M\\0&S\end{bmatrix},
    &\mathsf{G}_M&\coloneqq\begin{bmatrix}\mathsf B_M\\J\end{bmatrix},\\
    E&\coloneqq\begin{bmatrix}E_{\mathrm{o}}\\0_{Lm\times p}\end{bmatrix},
    &C&\coloneqq[\,C_{\mathrm{o}}\quad0_{p\times Lm}\,],
\end{aligned}
\]
where \(\mathsf{F}_M\in\mathbb R^{n\times n}\),
\(\mathsf{G}_M\in\mathbb R^{n\times m}\), \(E\in\mathbb R^{n\times p}\),
and \(C\in\mathbb R^{p\times n}\). Substituting \(M=\Theta\) gives
the true history dynamics from \eqref{eq:system} and, in particular,
\(\xi_{k+1}=\mathsf{F}_\Theta\xi_k+\mathsf{G}_\Theta v_k+Ew_k\).
At an online update, the output measurements and stored applied
references supply the initial history.

\begin{proposition}[Certified prediction tubes]
\label{prop:tube}

Let $\mathcal P$ in Proposition~\ref{prop:data} be compact.
Fix the model set, a feasible nominal $\widehat\Theta\in\mathcal P$,
and a vector norm $\|\cdot\|_{\mathrm{o}}$ on $\mathbb R^{n_{\mathrm{o}}}$.
The subscript \(\mathrm{o}\) denotes the output-history space.
The associated induced matrix norm is
\[
    \|A\|_{\mathrm{o}}
    \coloneqq\sup_{x\ne0}\frac{\|Ax\|_{\mathrm{o}}}{\|x\|_{\mathrm{o}}},
    \qquad A\in\mathbb R^{n_{\mathrm{o}}\times n_{\mathrm{o}}},
\]
where \(x\in\mathbb R^{n_{\mathrm{o}}}\).
The benchmark in \eqref{num:norms} uses a weighted infinity norm as
one example. Evaluate the candidate maps at \(\widehat\Theta\) to
obtain the nominal maps
\[
    \widehat{\mathsf A}=\mathsf A_{\widehat\Theta},
    \qquad
    \widehat{\mathsf D}=\mathsf D_{\widehat\Theta},
    \qquad
    \widehat{\mathsf B}=\mathsf B_{\widehat\Theta},
\]
\[
    \widehat{\mathsf{F}}=\mathsf{F}_{\widehat\Theta},
    \qquad
    \widehat{\mathsf{G}}=\mathsf{G}_{\widehat\Theta}.
\]
To bound how an output-history error propagates, add the nominal gain
\(a\) and the model deviation allowance \(\delta_A\) to obtain \(q\).
The same norm measures the injected disturbance in the definition of
\(\omega\). These quantities are
\[
\begin{aligned}
    a&\coloneqq\|\widehat{\mathsf A}\|_{\mathrm{o}},
    &\delta_A&\coloneqq\max_{M\in\mathcal P}
        \|\mathsf A_M-\widehat{\mathsf A}\|_{\mathrm{o}},\\
    q&\coloneqq a+\delta_A,
    &\omega&\coloneqq\max_{w\in\mathcal W}\|E_{\mathrm{o}}w\|_{\mathrm{o}},
\end{aligned}
\]
At a specified history \(\xi=\operatorname{col}(z,u)\) and reference
\(v\), the one-step model mismatch in the output-history norm is
bounded by
\begin{equation}
\begin{split}
    \beta(\xi,v)
    \coloneqq\max_{M\in\mathcal P}\bigl\|
        &(\mathsf A_M-\widehat{\mathsf A})z
        +(\mathsf D_M-\widehat{\mathsf D})u\\
        &+(\mathsf B_M-\widehat{\mathsf B})v\bigr\|_{\mathrm{o}}.
\end{split}
    \label{eq:lifted-uncertainty}
\end{equation}
Although the applied-reference history is exact, the candidate
coefficients on past and current references contribute to \(\beta\)
along with those on the output history. The defining maxima are finite
and attained, and $\beta$ is continuous and convex in $(\xi,v)$.

At a fixed online time \(k\), step \(j\) is relative to the initial
history and corresponds to time \(k+j\). For an initial history
$\xi=\operatorname{col}(z,u)$ and a reference $v$ held constant during
prediction, define the nominal center by
\begin{equation}
\begin{aligned}
    \widehat z_0&=z, & u_0&=u,\\
    \widehat z_{j+1}
        &=\widehat{\mathsf A}\widehat z_j
          +\widehat{\mathsf D}u_j+\widehat{\mathsf B}v,
    &u_{j+1}&=Su_j+Jv,\\
    \widehat\xi_j&=\operatorname{col}(\widehat z_j,u_j),
\end{aligned}
\label{eq:nominal-prediction}
\end{equation}
The radius accounts for propagation of the current output-history error,
one-step model mismatch at the nominal center, and disturbance. It obeys
\begin{equation}
    \rho_0=0,
    \qquad
    \rho_{j+1}=q\rho_j+\beta(\widehat\xi_j,v)+\omega,
    \label{eq:tube-recursion}
\end{equation}
The prediction tube at each step consists of all histories in this
output-history ball with the exact reference block. Define it by
\begin{equation}
    Z_j(\xi,v)
    \coloneqq\left\{
        \operatorname{col}(\widehat z_j+e,u_j):
        \|e\|_{\mathrm{o}}\leq\rho_j
      \right\}.
    \label{eq:prediction-tube}
\end{equation}
The following properties hold.

\begin{enumerate}
    \item \textit{Trajectory containment.}
    Every trajectory initialized at $\xi$, with a fixed
    $M\in\mathcal P$, a held reference $v$, and disturbances
    in $\mathcal W$, belongs to $Z_j(\xi,v)$ at every step
    $j\geq0$.

    \item \textit{Whole-tube inclusion after restart.}
    For every successor
    \[
        \zeta=\mathsf{F}_M\xi+\mathsf{G}_Mv+Ew,
        \qquad M\in\mathcal P,\quad w\in\mathcal W,
    \]
    restart the predictions from $\zeta$ using the same model set,
    nominal, norm, and held reference. The entire restarted tube satisfies
    \begin{equation}
        Z_j(\zeta,v)\subseteq Z_{j+1}(\xi,v),
        \qquad j\geq0.
        \label{eq:tube-reset}
    \end{equation}

    \item \textit{Bounded tail under contraction.}
    If $q<1$, use the nominal equilibrium under the held reference.
    Let $u_{\mathrm{e}}(v)=\operatorname{col}(v,\ldots,v)$ contain $L$
    copies of $v$, and define
    \[
    \begin{gathered}
        \widehat z_{\mathrm{e}}(v)
        =(I_{n_{\mathrm{o}}}-\widehat{\mathsf A})^{-1}
          (\widehat{\mathsf D}u_{\mathrm{e}}(v)+\widehat{\mathsf B}v),\\
        \widehat\xi_{\mathrm{e}}(v)
        =\operatorname{col}(\widehat z_{\mathrm{e}}(v),u_{\mathrm{e}}(v)),
    \end{gathered}
    \]
    \[
        \eta(v)=\beta(\widehat\xi_{\mathrm{e}}(v),v).
    \]
    The vector $u_{\mathrm{e}}(v)$ is empty for $L=0$.
    For every integer $t\geq1$,
    \begin{equation}
    \begin{split}
        \rho_{L+t}\leq{}&q^t\rho_L
        +\delta_A\|\widehat z_L-\widehat z_{\mathrm{e}}(v)\|_{\mathrm{o}}
          \sum_{s=0}^{t-1}q^{t-1-s}a^s\\
        &+(\eta(v)+\omega)\frac{1-q^t}{1-q}.
    \end{split}
    \label{eq:tube-tail}
    \end{equation}
    Consequently,
    \[
        \limsup_{j\to\infty}\rho_j
        \leq\frac{\eta(v)+\omega}{1-q}.
    \]
\end{enumerate}
The first two conclusions hold for every $q\geq0$.

\end{proposition}

The terminal construction below uses \(q<1\) as a sufficient
contraction condition. Asymptotic stability of the true system
can hold even when \(q\geq1\). The test evaluates the candidate
operators in the chosen history norm independently of any
physical-output overshoot.

\begin{proof}
\textit{Uniform bounds and regularity.}
The lifted matrices depend affinely on $M$. Compactness of
$\mathcal P$ and $\mathcal W$ therefore makes all defining
maxima finite and attained. For every $M\in\mathcal P$,
\[
    \|\mathsf A_M\|_{\mathrm{o}}
    \leq\|\widehat{\mathsf A}\|_{\mathrm{o}}
         +\|\mathsf A_M-\widehat{\mathsf A}\|_{\mathrm{o}}
    \leq q.
\]
Each function maximized in $\beta$ is the norm of a linear
map of $(\xi,v)$, hence convex. Compactness gives a uniform
bound on these linear maps, so their maximum is continuous.
For histories with the same reference block, the triangle
inequality in both directions gives
\begin{equation}
    \left|
        \beta(\operatorname{col}(z,u),v)
        -\beta(\operatorname{col}(z',u),v)
    \right|
    \leq\delta_A\|z-z'\|_{\mathrm{o}}.
    \label{eq:beta-lipschitz}
\end{equation}

\textit{Trajectory containment.}
The reference-history recursion is exact. For a fixed
$M\in\mathcal P$, write \(z_j^M\) for its output-history trajectory
and set $e_j\coloneqq z_j^M-\widehat z_j$.
The shared initial history gives \(e_0=0\).
Subtracting \eqref{eq:nominal-prediction} from the candidate update
\eqref{eq:candidate-history-dynamics} gives
\[
\begin{split}
    e_{j+1}={}&\mathsf A_Me_j
       +(\mathsf A_M-\widehat{\mathsf A})\widehat z_j
       +(\mathsf D_M-\widehat{\mathsf D})u_j\\
       &+(\mathsf B_M-\widehat{\mathsf B})v+E_{\mathrm{o}}w_j.
\end{split}
\]
Thus
\[
    \|e_{j+1}\|_{\mathrm{o}}
    \leq q\|e_j\|_{\mathrm{o}}+\beta(\widehat\xi_j,v)+\omega.
\]
Comparison with \eqref{eq:tube-recursion} proves
$\|e_j\|_{\mathrm{o}}\leq\rho_j$ by induction. In particular, the
true trajectory is contained because
Proposition~\ref{prop:data} gives $\Theta\in\mathcal P$.

\textit{Reset inclusion.}
Write $\zeta=\operatorname{col}(z^+,u^+)$, and use primes for
predictions recomputed from that successor history. Align step
\(j\) of the restarted prediction with step \(j+1\) of the original. The first-step bound
gives $\|z^+-\widehat z_1\|_{\mathrm{o}}\leq\rho_1$.
The held reference gives
\[
    u'_j=u_{j+1},
    \qquad
    \widehat z'_j-\widehat z_{j+1}
    =\widehat{\mathsf A}^{\,j}(z^+-\widehat z_1).
\]
Let $b_0=\rho_1$ and $b_{j+1}=ab_j$. The center difference
is at most $b_j$. The sum of this displacement bound and the
restarted radius bounds the distance from any point of the restarted
tube to the original output-history center. Starting from
$\rho'_0+b_0=\rho_1$, suppose
$\rho'_j+b_j\leq\rho_{j+1}$. The equal reference blocks
permit application of \eqref{eq:beta-lipschitz}, giving
\begin{align*}
    \rho'_{j+1}+b_{j+1}
    &\leq q\rho'_j+\beta(\widehat\xi_{j+1},v)
           +\delta_Ab_j+\omega+ab_j\\
    &=q(\rho'_j+b_j)+\beta(\widehat\xi_{j+1},v)+\omega\\
    &\leq\rho_{j+2}.
\end{align*}
Every point of the restarted tube therefore has the same
reference block as the old tube and is within
$\rho'_j+b_j\leq\rho_{j+1}$ of its output-history center.
This proves \eqref{eq:tube-reset} for the entire tube.

\textit{Bounded tail.}
If $q<1$, then $a\leq q<1$, so
$I_{n_{\mathrm{o}}}-\widehat{\mathsf A}$ is invertible.
After $L$ held-reference steps, $u_j=u_{\mathrm{e}}(v)$.
Consequently,
\[
    \widehat z_{L+s}-\widehat z_{\mathrm{e}}(v)
    =\widehat{\mathsf A}^{\,s}
       (\widehat z_L-\widehat z_{\mathrm{e}}(v)),
    \qquad s\geq0.
\]
Using \eqref{eq:beta-lipschitz} on this common reference block,
\[
    \beta(\widehat\xi_{L+s},v)
    \leq\eta(v)
       +\delta_Aa^s\|\widehat z_L-\widehat z_{\mathrm{e}}(v)\|_{\mathrm{o}}.
\]
Expanding \eqref{eq:tube-recursion} from $\rho_L$ gives
\eqref{eq:tube-tail}. For $0<q<1$, its transient sum is
bounded by $tq^{t-1}$ and tends to zero. If $q=0$, then
$a=\delta_A=0$ and the recursion reaches
$\eta(v)+\omega$ after one further step. The stated limiting
bound follows in both cases.
\end{proof}

The identity
\[
    (\mathsf A_M-\widehat{\mathsf A})z
    +(\mathsf D_M-\widehat{\mathsf D})u
    +(\mathsf B_M-\widehat{\mathsf B})v
    =E_{\mathrm{o}}(M-\widehat\Theta)\phi(\xi,v)
\]
connects $\beta$ to the directional uncertainty in
Proposition~\ref{prop:data}. For the unweighted infinity
norm on output histories,
$\beta(\xi,v)=\epsilon(\phi(\xi,v))$ and $\omega=\bar w$.

The containment argument also permits a different compatible
matrix at each prediction step. The fixed-model envelope in
the next proposition separates uncertainty retained in
$\mathcal P$ from the additional overestimation introduced
by this propagation.

\subsection{Constraint margins}

At each prediction step, $C_{\mathrm{o}}$ reads the output from the
output-history block. Applying the constraint row $H_i$ gives the
corresponding scalar value. The support of the output-history error
ball in that direction gives the margin above the nominal row value.
To express the margin per unit radius, define the dual norm and the
row coefficient by
\[
    \|\lambda\|_{\mathrm{o},*}
    \coloneqq\max_{\|e\|_{\mathrm{o}}\leq1}|\lambda e|,
    \qquad
    d_i\coloneqq\|H_iC_{\mathrm{o}}\|_{\mathrm{o},*},
\]
where $\lambda\in\mathbb R^{1\times n_{\mathrm{o}}}$ is a row vector
and $e\in\mathbb R^{n_{\mathrm{o}}}$ is an output-history error.
The margin bound uses finite-horizon containment from
Proposition~\ref{prop:tube} for any $q\geq0$.

\begin{proposition}[Constraint margins and conservatism]
\label{prop:margins}

Use the compact model set and prediction tubes from
Proposition~\ref{prop:tube}. Fix an initial history
$\xi=\operatorname{col}(z,u)$ and a reference
$v\in\mathcal V$ held constant during prediction.
The following conclusions apply to each constraint row $i$ and
relative prediction step $j\geq0$.

\begin{enumerate}
    \item \textit{Tube margin and row test.}
The margin and the resulting tube envelope are
\begin{equation}
    \Delta_{i,j}\coloneqq d_i\rho_j,
    \qquad
    b_{i,j}^{\mathrm{tube}}
    \coloneqq H_iC_{\mathrm{o}}\widehat z_j+\Delta_{i,j}.
    \label{eq:constraint-margin}
\end{equation}
Every compatible trajectory satisfies
\begin{equation}
    H_i y_{k+j}
    \leq\max_{\zeta\in Z_j(\xi,v)}H_iC\zeta
    =b_{i,j}^{\mathrm{tube}}.
    \label{eq:constraint-envelope}
\end{equation}
Thus the test
\begin{equation}
    b_{i,j}^{\mathrm{tube}}\leq h_i
    \label{eq:tightened-constraint}
\end{equation}
certifies constraint $i$ at step $j$.

    \item \textit{Uncertainty utilization.}
The unused nominal clearance is
$c_{i,j}\coloneqq h_i-H_iC_{\mathrm{o}}\widehat z_j$. When it
is positive, the fraction consumed by the margin is
\begin{equation}
    U_{i,j}\coloneqq\frac{\Delta_{i,j}}{c_{i,j}}.
    \label{eq:uncertainty-utilization}
\end{equation}
Then $U_{i,j}\leq1$ is equivalent to
\eqref{eq:tightened-constraint}, and the ratio is unchanged
by positive rescaling of $(H_i,h_i)$.
If $c_{i,j}=0$, the tube test holds exactly when
$\Delta_{i,j}=0$; if $c_{i,j}<0$, it fails.

    \item \textit{Fixed-model envelope.}
To compare the tube with the outputs reachable under one model,
consider every $M\in\mathcal P$, holding it fixed throughout its rollout.
Use the same deterministic reference-history sequence
$u_t$ as in \eqref{eq:nominal-prediction}. The candidate dynamics give
\[
    z_0^M=z,
    \qquad
    z_{t+1}^M
    =\mathsf A_Mz_t^M+\mathsf D_Mu_t+\mathsf B_Mv+E_{\mathrm{o}}w_t.
\]
The envelope maximizes the resulting row value over disturbance sequences,
\[
    b_{i,j}(M)
    \coloneqq\max_{(w_0,\ldots,w_{j-1})\in\mathcal W^j}
        H_iC_{\mathrm{o}}z_j^M.
\]
The support of the disturbance box in \eqref{eq:disturbance}
gives the expression
\begin{equation}
\begin{split}
    b_{i,j}(M)={}&
    H_iC_{\mathrm{o}}\left(
        \mathsf A_M^jz+
        \sum_{t=0}^{j-1}\mathsf A_M^{j-1-t}
          (\mathsf D_Mu_t+\mathsf B_Mv)
    \right)\\
    &+\bar w\sum_{t=0}^{j-1}
          \|H_iC_{\mathrm{o}}\mathsf A_M^tE_{\mathrm{o}}\|_1.
\end{split}
\label{eq:fixed-model-bound}
\end{equation}
At $j=0$, the sums are empty and $b_{i,0}(M)=H_iC_{\mathrm{o}}z$.

    \item \textit{Ordered bounds and losses.}
The true-model envelope uses $\Theta$. The data-compatible
envelope maximizes over the compatible fixed models. Both use the
same initial history, held reference, and disturbance set as the tube,
\[
    b_{i,j}^{\mathrm{true}}\coloneqq b_{i,j}(\Theta),
    \qquad
    b_{i,j}^{\mathrm{data}}
    \coloneqq\max_{M\in\mathcal P}b_{i,j}(M).
\]
For every finite $j$, the defining maxima are finite and attained,
and the envelopes satisfy
\begin{equation}
    b_{i,j}^{\mathrm{true}}
    \leq b_{i,j}^{\mathrm{data}}
    \leq b_{i,j}^{\mathrm{tube}}.
    \label{eq:ordered-bounds}
\end{equation}
The information loss compares the true-model and data-compatible
envelopes. The propagation loss measures the additional gap from the
tube geometry and its uniform propagation bounds. The differences
\begin{align}
    L_{i,j}^{\mathrm{information}}
    &\coloneqq b_{i,j}^{\mathrm{data}}-b_{i,j}^{\mathrm{true}},
    \label{eq:information-loss}\\
    L_{i,j}^{\mathrm{propagation}}
    &\coloneqq b_{i,j}^{\mathrm{tube}}-b_{i,j}^{\mathrm{data}}
    \label{eq:propagation-loss}
\end{align}
are nonnegative and satisfy
\begin{equation}
    b_{i,j}^{\mathrm{tube}}-b_{i,j}^{\mathrm{true}}
    =L_{i,j}^{\mathrm{information}}
     +L_{i,j}^{\mathrm{propagation}}.
    \label{eq:conservatism-decomposition}
\end{equation}
At $j=0$, all three bounds coincide and both losses are zero.
\end{enumerate}
\end{proposition}

\begin{proof}
\textit{Constraint margin.}
Trajectory containment gives $\xi_{k+j}\in Z_j(\xi,v)$.
The exact reference block contributes no output error, so
\begin{align*}
    \max_{\zeta\in Z_j(\xi,v)}H_iC\zeta
    &=H_iC_{\mathrm{o}}\widehat z_j
      +\max_{\|e\|_{\mathrm{o}}\leq\rho_j}H_iC_{\mathrm{o}}e\\
    &=H_iC_{\mathrm{o}}\widehat z_j+d_i\rho_j.
\end{align*}
The continuous row functional attains its maximum on the compact
norm ball. Scaling the unit ball and using its symmetry give the
equality for $\rho_j>0$. For $\rho_j=0$, the ball contains only zero.
This proves \eqref{eq:constraint-envelope} and the sufficient
test \eqref{eq:tightened-constraint}.

\textit{Uncertainty utilization.}
The tube test is equivalent to $\Delta_{i,j}\leq c_{i,j}$.
Division by $c_{i,j}>0$ gives the stated equivalence.
Multiplying $(H_i,h_i)$ by a positive scalar multiplies
$d_i$, $\Delta_{i,j}$, and $c_{i,j}$ by that scalar, leaving
$U_{i,j}$ unchanged. The cases of zero and negative clearance
follow from $\Delta_{i,j}\geq0$.

\textit{Fixed-model envelope.}
For a fixed $M$, expand the output-history recursion to isolate
the contribution of each disturbance,
\[
    z_j^M=\mathsf A_M^jz+
    \sum_{t=0}^{j-1}\mathsf A_M^{j-1-t}
       (\mathsf D_Mu_t+\mathsf B_Mv+E_{\mathrm{o}}w_t).
\]
For any row $\lambda\in\mathbb R^{1\times p}$,
\[
    \max_{\|w\|_\infty\leq\bar w}\lambda w
    =\bar w\|\lambda\|_1,
\]
with a maximizer obtained by choosing each component of $w$
with the sign of the corresponding component of $\lambda$.
The disturbance sequences range over the product
$\mathcal W^j$, so their separate linear contributions can
be maximized independently. Applying this support formula
and reindexing the disturbance sum gives
\eqref{eq:fixed-model-bound}.

For fixed $M$ and finite $j$, the disturbance maximum is
attained because $\mathcal W^j$ is compact and the objective
is continuous. The reference sequence $u_t$ is independent
of $M$. Since $\mathsf A_M$, $\mathsf D_M$, and
$\mathsf B_M$ depend affinely on $M$,
\eqref{eq:fixed-model-bound} is continuous in $M$.
Compactness of $\mathcal P$ therefore gives a finite attained
maximum for $b_{i,j}^{\mathrm{data}}$. The case $j=0$ follows
directly from the common initial history.

\textit{Ordered bounds and losses.}
True-model containment gives
$b_{i,j}^{\mathrm{true}}\leq b_{i,j}^{\mathrm{data}}$.
For every fixed $M\in\mathcal P$ and disturbance sequence,
Proposition~\ref{prop:tube} gives
\[
    H_iC_{\mathrm{o}}z_j^M\leq b_{i,j}^{\mathrm{tube}}.
\]
Maximizing over disturbances and then over $M$ proves
$b_{i,j}^{\mathrm{data}}\leq b_{i,j}^{\mathrm{tube}}$.
The losses are therefore nonnegative, and their sum is
\eqref{eq:conservatism-decomposition}.
\end{proof}

The tube test is necessary and sufficient for the constraint to
hold throughout the chosen tube, so passing it certifies the row
for every compatible trajectory. A failed test can result from a
violating tube point outside all compatible fixed-model rollouts.

Once the uncertainty evaluations required by
\eqref{eq:tube-recursion} are available, the DDRG computes
$b_{i,j}^{\mathrm{tube}}$, the row slack $h_i-b_{i,j}^{\mathrm{tube}}$,
and the positive-clearance
utilization online. The data-compatible envelope is a comparison
quantity whose computation may require nonconvex optimization over
$M$. The true-model envelope is an evaluation quantity when the plant
is known.
For fixed $\mathcal P$, the data-compatible envelope is
independent of the nominal model and tube norm, whereas the
propagation loss depends on those design choices.

Uncertainty utilization diagnoses the stage constraints.
Proposition~\ref{prop:admissibility} adds two tests for a certified
reference update. The reference must admit a safe terminal region,
and the predicted tube must enter that region.

\subsection{Invariant admissibility}

The stage tests bound the output through a finite rollout under a held
reference. For continuation, the candidate history update preserves the
equilibrium reference history. Under the contraction condition, the
output-history deviation from equilibrium is bounded by a fraction of
the previous deviation plus mismatch and disturbance terms.
A terminal fiber uses this bound to certify safe continuation of
the entire horizon tube, including after a restart from the next
measured history.

\begin{proposition}[Invariant admissibility certificate]
\label{prop:admissibility}
Use the model set, nominal model, output-history norm, and prediction
tubes of Propositions~\ref{prop:data}--\ref{prop:margins}.
Keep these choices fixed and suppose that $q<1$.
Choose an integer $N_{\mathrm{p}}\geq\max\{1,L\}$, where $L=\ell-1$.
The condition $N_{\mathrm{p}}\geq L$ ensures that the held reference
has filled the stored reference history by the horizon.

For a reference held at $v$, the terminal fiber is centered at the
nominal equilibrium. For this certificate, the radius must absorb
equilibrium mismatch and disturbances and accommodate the entering tube.
The first requirement sets a radius floor $\mu(v)$. The second uses
the sum $\tau_j(\xi,v)$ of the distance from the tube center to
equilibrium and the tube radius at step $j$. These quantities are
\begin{equation}
    \mu(v)\coloneqq\frac{\eta(v)+\omega}{1-q},
    \qquad
    \tau_j(\xi,v)
    \coloneqq\|\widehat z_j-\widehat z_{\mathrm{e}}(v)\|_{\mathrm{o}}+\rho_j,
    \label{eq:terminal-floor-entry}
\end{equation}
where the nominal predictions and radii are initialized at $\xi$.
The equilibrium output leaves a rowwise margin for each of these
requirements. The remaining slacks for the floor and entry tests are
\begin{equation}
\begin{split}
    s_i^{\mathrm{floor}}(v)
    &\coloneqq h_i-H_iC_{\mathrm{o}}\widehat z_{\mathrm{e}}(v)-d_i\mu(v),\\
    s_i^{\mathrm{entry}}(\xi,v)
    &\coloneqq h_i-H_iC_{\mathrm{o}}\widehat z_{\mathrm{e}}(v)-d_i\tau_{N_{\mathrm{p}}}(\xi,v).
\end{split}
    \label{eq:terminal-slacks}
\end{equation}

The references that leave enough margin for the radius floor form
$\mathcal V_*$, the set of $v\in\mathcal V$ satisfying
\begin{equation}
    H_iC_{\mathrm{o}}\widehat z_{\mathrm{e}}(v)+d_i\mu(v)\leq h_i,
    \qquad i=1,\ldots,n_c.
    \label{eq:terminal-references}
\end{equation}
\begin{samepage}
The admissible set $\mathcal S_{N_{\mathrm{p}}}$ consists of pairs
$(\xi,v)$ with $v\in\mathcal V$ that pass the stage tests before the
horizon, the radius-floor test, and the tube-entry test at the horizon.
For each row $i=1,\ldots,n_c$, the defining inequalities are
\begin{equation}
\begin{aligned}
    H_iC_{\mathrm{o}}\widehat z_j+d_i\rho_j&\leq h_i,
    &&j=0,\ldots,N_{\mathrm{p}}-1,\\
    H_iC_{\mathrm{o}}\widehat z_{\mathrm{e}}(v)+d_i\mu(v)&\leq h_i,\\
    H_iC_{\mathrm{o}}\widehat z_{\mathrm{e}}(v)+d_i\tau_{N_{\mathrm{p}}}(\xi,v)&\leq h_i.
\end{aligned}
    \label{eq:finite-horizon-admissibility}
\end{equation}
\end{samepage}

To show why these tests allow a safe continuation, fix a radius
$R\geq0$. The reference set $\mathcal V_R$ consists of the
$v\in\mathcal V$ for which $R$ covers the floor and keeps the
equilibrium tube within each constraint limit,
\begin{equation}
    \mu(v)\leq R,
    \qquad H_iC_{\mathrm{o}}\widehat z_{\mathrm{e}}(v)+d_iR\leq h_i,
    \quad i=1,\ldots,n_c.
    \label{eq:fixed-radius-references}
\end{equation}
The terminal fiber combines the output-history ball with the exact
equilibrium reference history,
\begin{equation}
    \mathcal X_R(v)
    \coloneqq
    \left\{
        \operatorname{col}(z,u_{\mathrm{e}}(v)):
        \|z-\widehat z_{\mathrm{e}}(v)\|_{\mathrm{o}}\leq R
    \right\}.
    \label{eq:terminal-set}
\end{equation}
The fixed-radius admissible set $\mathcal S_{N_{\mathrm{p}}}(R)$
consists of pairs $(\xi,v)$ with $v\in\mathcal V_R$ that pass the
same stage tests and whose horizon tube fits in this fiber. These pairs
satisfy
\begin{equation}
\begin{aligned}
    H_iC_{\mathrm{o}}\widehat z_j+d_i\rho_j&\leq h_i,
    &&i=1,\ldots,n_c,\quad j=0,\ldots,N_{\mathrm{p}}-1,\\
    \tau_{N_{\mathrm{p}}}(\xi,v)&\leq R.
\end{aligned}
    \label{eq:fixed-radius-admissibility}
\end{equation}

\begin{enumerate}
    \item \textit{Radius elimination.}
The fixed-radius sets recover the direct tests through
\begin{equation}
    \mathcal S_{N_{\mathrm{p}}}
    =\bigcup_{R\geq0}\mathcal S_{N_{\mathrm{p}}}(R),
    \qquad
    \mathcal V_*=\bigcup_{R\geq0}\mathcal V_R.
    \label{eq:radius-elimination}
\end{equation}
For any $(\xi,v)\in\mathcal S_{N_{\mathrm{p}}}$, a feasible radius is
\begin{equation}
    R_*(\xi,v)\coloneqq\max\{\mu(v),\tau_{N_{\mathrm{p}}}(\xi,v)\}.
    \label{eq:terminal-radius-witness}
\end{equation}
Evaluate the certificate directly using
\eqref{eq:finite-horizon-admissibility}. The radius $R_*$ serves as a proof
witness and need not be stored.

    \item \textit{Safety and invariance.}
For $v\in\mathcal V_R$, the terminal fiber $\mathcal X_R(v)$
satisfies the output constraints and is robustly invariant under
the held reference. Every prediction tube that enters this fiber
remains in it. Both $\mathcal S_{N_{\mathrm{p}}}(R)$ and $\mathcal S_{N_{\mathrm{p}}}$
are closed and robustly invariant under the held reference. In
particular,
\begin{equation}
\begin{split}
    (\xi,v)\in\mathcal S_{N_{\mathrm{p}}}
    \quad\Longrightarrow\quad
    &(\mathsf{F}_M\xi+\mathsf{G}_Mv+Ew,v)\in\mathcal S_{N_{\mathrm{p}}},\\
    &\forall M\in\mathcal P,\quad \forall w\in\mathcal W.
\end{split}
    \label{eq:admissible-set-invariance}
\end{equation}
Consequently, a pair in either admissible set satisfies the
constraints indefinitely while its reference is held.
    \item \textit{Horizon nesting and set properties.}
The admissible sets satisfy
\begin{equation}
    \mathcal S_{N_{\mathrm{p}}}(R)\subseteq\mathcal S_{N_{\mathrm{p}}+1}(R),
    \qquad
    \mathcal S_{N_{\mathrm{p}}}\subseteq\mathcal S_{N_{\mathrm{p}}+1}.
    \label{eq:horizon-nesting}
\end{equation}
The set $\mathcal S_{N_{\mathrm{p}}}$ is nonempty exactly when
$\mathcal V_*$ is nonempty. The sets $\mathcal V_R$ and
$\mathcal V_*$ are compact. If $\mathcal V$ is convex, all four
sets $\mathcal V_R$, $\mathcal V_*$, $\mathcal S_{N_{\mathrm{p}}}(R)$,
and $\mathcal S_{N_{\mathrm{p}}}$ are convex.
\end{enumerate}
\end{proposition}

\begin{proof}
\textit{Safe invariant fiber.}
Fix $v\in\mathcal V_R$ and a history
$\xi=\operatorname{col}(z,u_{\mathrm{e}}(v))\in\mathcal X_R(v)$.
The exact reference recursion preserves $u_{\mathrm{e}}(v)$.
For any $M\in\mathcal P$ and $w\in\mathcal W$, the output
block of the successor satisfies
\[
\begin{split}
    z^+-\widehat z_{\mathrm{e}}(v)
    ={}&\mathsf A_M(z-\widehat z_{\mathrm{e}}(v))
       +(\mathsf A_M-\widehat{\mathsf A})\widehat z_{\mathrm{e}}(v)\\
      &+(\mathsf D_M-\widehat{\mathsf D})u_{\mathrm{e}}(v)
       +(\mathsf B_M-\widehat{\mathsf B})v+E_{\mathrm{o}}w.
\end{split}
\]
Since $\|\mathsf A_M\|_{\mathrm{o}}\leq q$,
\[
\begin{aligned}
    \|z^+-\widehat z_{\mathrm{e}}(v)\|_{\mathrm{o}}
    &\leq q\|z-\widehat z_{\mathrm{e}}(v)\|_{\mathrm{o}}+\eta(v)+\omega\\
    &\leq qR+(1-q)R=R.
\end{aligned}
\]
The dual-norm inequality gives
\[
    H_iC_{\mathrm{o}}z
    \leq H_iC_{\mathrm{o}}\widehat z_{\mathrm{e}}(v)+d_iR\leq h_i.
\]
Thus the terminal fiber is safe and invariant, including when
$R=0$. In that case $\mu(v)\leq R$ forces
$\eta(v)+\omega=0$ and the terminal singleton is preserved.

\textit{Tube entry and continuation.}
To test a tube for inclusion in the terminal fiber, its exact
reference block must agree with the equilibrium reference history.
The held reference gives
$u_j=u_{\mathrm{e}}(v)$ for $j\geq L$. At each such stage,
\[
    \max_{\operatorname{col}(z',u_j)\in Z_j(\xi,v)}
    \|z'-\widehat z_{\mathrm{e}}(v)\|_{\mathrm{o}}
    =\|\widehat z_j-\widehat z_{\mathrm{e}}(v)\|_{\mathrm{o}}+\rho_j
    =\tau_j.
\]
The triangle inequality gives the upper bound. An error directed
along the center displacement attains it. If the displacement is
zero, any unit direction gives the equality.
Hence $Z_j(\xi,v)\subseteq\mathcal X_R(v)$ is equivalent to
$\tau_j\leq R$ once the reference history has flushed.
To continue the tube after entry, apply the nominal recursion and
the Lipschitz bound for $\beta$ at identical reference histories.
They give
\begin{align*}
    \tau_{j+1}
    &\leq a\|\widehat z_j-\widehat z_{\mathrm{e}}(v)\|_{\mathrm{o}}
         +q\rho_j+\beta(\widehat\xi_j,v)+\omega\\
    &\leq (a+\delta_A)\|\widehat z_j-\widehat z_{\mathrm{e}}(v)\|_{\mathrm{o}}
         +q\rho_j+\eta(v)+\omega\\
    &=q\tau_j+\eta(v)+\omega.
\end{align*}
Thus $\tau_j\leq R$ implies $\tau_{j+1}\leq R$.
For a prediction initialized in $\mathcal X_R(v)$, the reference
history already equals $u_{\mathrm{e}}(v)$, so this argument starts at
$j=0$. In particular,
$\mathcal X_R(v)\times\{v\}\subseteq\mathcal S_{N_{\mathrm{p}}}(R)$.

\textit{Restart and horizon extension.}
Let $(\xi,v)\in\mathcal S_{N_{\mathrm{p}}}(R)$ and let
$\zeta=\mathsf{F}_M\xi+\mathsf{G}_Mv+Ew$ be an admissible successor.
Proposition~\ref{prop:tube} gives
\[
    Z_j(\zeta,v)\subseteq Z_{j+1}(\xi,v),\qquad j\geq0.
\]
The original tubes at stages $1,\ldots,N_{\mathrm{p}}-1$ pass
the stage tests. The original tube at $N_{\mathrm{p}}$ lies in the
safe terminal fiber. Together they establish every restarted
stage test.
Terminal continuation also gives
\[
    Z_{N_{\mathrm{p}}}(\zeta,v)
    \subseteq Z_{N_{\mathrm{p}}+1}(\xi,v)
    \subseteq\mathcal X_R(v).
\]
This proves fixed-radius hold invariance. The old terminal tube
is also the additional safe stage for horizon $N_{\mathrm{p}}+1$, and its
successor remains in the same fiber. This proves fixed-radius
horizon nesting.

\textit{Radius elimination.}
For a fixed-radius feasible pair, $\mu(v),\tau_{N_{\mathrm{p}}}(\xi,v)\leq R$.
Since $d_i\geq0$, the pair satisfies both terminal inequalities
in \eqref{eq:finite-horizon-admissibility}.
Conversely, those inequalities imply
\[
    H_iC_{\mathrm{o}}\widehat z_{\mathrm{e}}(v)+d_iR_*(\xi,v)\leq h_i
\]
for every row. The finite nonnegative radius in
\eqref{eq:terminal-radius-witness} consequently satisfies all
fixed-radius tests. This proves the first equality in
\eqref{eq:radius-elimination}. The second follows by taking
$R=\mu(v)$ for $v\in\mathcal V_*$. No division by $d_i$ is
used, so zero constraint rows are included.
Fixed-radius invariance certifies each successor with the same
witness $R_*$, proving \eqref{eq:admissible-set-invariance}.
Taking unions in fixed-radius horizon nesting proves
\eqref{eq:horizon-nesting} for $\mathcal S_{N_{\mathrm{p}}}$.
The step-zero test enforces the current output constraints,
and invariance enforces them at all subsequent samples.

\textit{Nonemptiness.}
For $v\in\mathcal V_*$, the pair
$(\widehat\xi_{\mathrm{e}}(v),v)$ belongs to
$\mathcal S_{N_{\mathrm{p}}}(\mu(v))$. Conversely, membership in
$\mathcal S_{N_{\mathrm{p}}}$ implies $v\in\mathcal V_*$.
This establishes the nonemptiness assertion.

\textit{Closedness, compactness, and convexity.}
The function $\beta$ is continuous and convex, the nominal
predictions and equilibrium are linear, and $q\geq0$ is fixed.
The radius recursion therefore makes every $\rho_j$
continuous and convex in $(\xi,v)$. The same properties hold
for $\mu(v)$ and $\tau_{N_{\mathrm{p}}}(\xi,v)$.
The defining inequalities are continuous and $\mathcal V$ is
closed, which proves closedness directly for both admissible sets.
The reference sets are compact because they are closed subsets
of compact $\mathcal V$.
When $\mathcal V$ is convex, the defining inequalities are
convex, which proves the remaining assertions.
\end{proof}

Each row with a positive margin coefficient sets an upper bound
on the terminal radius. Define the most restrictive such bound by
\[
    R_{\max}(v)
    \coloneqq\min_{i:d_i>0}
      \frac{h_i-H_iC_{\mathrm{o}}\widehat z_{\mathrm{e}}(v)}{d_i},
\]
with the minimum over an empty index set equal to $+\infty$.
Rows with $d_i=0$ retain their direct inequalities
$H_iC_{\mathrm{o}}\widehat z_{\mathrm{e}}(v)\leq h_i$.
The terminal tests then read
$\mu(v)\leq R_{\max}(v)$ and
$\tau_{N_{\mathrm{p}}}(\xi,v)\leq R_{\max}(v)$, together with those
zero-row inequalities.
For $v\in\mathcal V_*$, continuation yields
\[
    \tau_{L+t}
    \leq q^t\tau_L+\mu(v)(1-q^t),\qquad t\geq0.
\]
If $R_{\max}(v)>\mu(v)$, terminal entry occurs at a finite
horizon whenever the required preceding stage tests hold.
At $R_{\max}(v)=\mu(v)$, this bound can approach the floor
without reaching it in finite time.
The reference set $\mathcal V_*$ is independent of the horizon;
increasing $N_{\mathrm{p}}$ enlarges admissible histories within that
reference set.

\subsection{Matched comparison with full-history balls}
The candidate history dynamics propagate the applied-reference
block exactly. The comparison assesses the admissible sets obtained
when the tube uses an error ball in the entire history. Both
certificates use the same nominal centers because they use the same
history dynamics and nominal model. Keep the data, prior restrictions,
model set, nominal model, disturbance set, constraints, and horizon
fixed.

To compare radii and constraint margins, choose a norm
$\|\cdot\|_{\mathrm f}$ on $\mathbb R^n$ whose restriction to the
zero-reference-error subspace is the selected output-history norm. Assume that deleting the
reference-history coordinates is nonexpansive,
\begin{equation}
    \|z\|_{\mathrm{o}}=\|\operatorname{col}(z,0)\|_{\mathrm f},
    \qquad
    \|\operatorname{col}(z,0)\|_{\mathrm f}
    \leq\|\operatorname{col}(z,u)\|_{\mathrm f}.
    \label{eq:matched-history-norms}
\end{equation}
For the full-history radius recursion, bound nominal propagation,
model variation, and disturbance injection with the induced
full-history matrix norm and its vector norm,
\[
\begin{aligned}
    a_{\mathrm f}&\coloneqq\|\widehat{\mathsf{F}}\|_{\mathrm f},
    &\delta_{\mathrm f}
       &\coloneqq\max_{M\in\mathcal P}\|\mathsf{F}_M-\widehat{\mathsf{F}}\|_{\mathrm f},\\
    q_{\mathrm f}&\coloneqq a_{\mathrm f}+\delta_{\mathrm f},
    &\omega_{\mathrm f}
       &\coloneqq\max_{w\in\mathcal W}\|Ew\|_{\mathrm f}.
\end{aligned}
\]
The local and equilibrium mismatches are
\[
\begin{aligned}
    \beta_{\mathrm f}(\xi,v)
       &\coloneqq\max_{M\in\mathcal P}
       \|(\mathsf{F}_M-\widehat{\mathsf{F}})\xi+(\mathsf{G}_M-\widehat{\mathsf{G}})v\|_{\mathrm f},\\
    \eta_{\mathrm f}(v)
       &\coloneqq\beta_{\mathrm f}(\widehat\xi_{\mathrm{e}}(v),v).
\end{aligned}
\]
Assume $q_{\mathrm f}<1$. The full-history radius is propagated
about the common nominal centers according to
\[
\begin{aligned}
    \rho^{\mathrm f}_0&=0,\\
    \rho^{\mathrm f}_{j+1}
    &=q_{\mathrm f}\rho^{\mathrm f}_j
      +\beta_{\mathrm f}(\widehat\xi_j,v)+\omega_{\mathrm f},\\
    Z_j^{\mathrm f}
    &\coloneqq\{\widehat\xi_j+e:\|e\|_{\mathrm f}\leq\rho_j^{\mathrm f}\}.
\end{aligned}
\]
Let $\|\cdot\|_{\mathrm f,*}$ denote the full-history dual norm
and $N$ the relative prediction index. The margin coefficient,
radius floor, and terminal distance bound are
\[
\begin{aligned}
    d_i^{\mathrm f}&\coloneqq\|H_iC\|_{\mathrm f,*},&
    \mu_{\mathrm f}(v)
       &\coloneqq\frac{\eta_{\mathrm f}(v)+\omega_{\mathrm f}}{1-q_{\mathrm f}},\\
    \tau_N^{\mathrm f}
       &\coloneqq\|\widehat\xi_N-\widehat\xi_{\mathrm{e}}(v)\|_{\mathrm f}
        +\rho_N^{\mathrm f}.
\end{aligned}
\]

For $R\geq0$, the full-history certificate
$\mathcal S_{N_{\mathrm{p}}}^{\mathrm f}(R)$ consists of the pairs
$(\xi,v)$ with $v\in\mathcal V$, $\mu_{\mathrm f}(v)\leq R$, and
$\tau_{N_{\mathrm{p}}}^{\mathrm f}(\xi,v)\leq R$ that satisfy, for
each $i=1,\ldots,n_c$,
\[
\begin{aligned}
    H_iC\widehat\xi_j+d_i^{\mathrm f}\rho_j^{\mathrm f}
        &\leq h_i, &&j=0,\ldots,N_{\mathrm{p}}-1,\\
    H_iC\widehat\xi_{\mathrm{e}}(v)+d_i^{\mathrm f}R&\leq h_i.
\end{aligned}
\]
Let $\mathcal S_{N_{\mathrm{p}}}^{\mathrm f}
\coloneqq\bigcup_{R\geq0}\mathcal S_{N_{\mathrm{p}}}^{\mathrm f}(R)$.
Equivalently, for membership in this union, keep the reference
membership and stage conditions and replace the three radius
conditions by the two terminal tests for each $i$,
\[
\begin{aligned}
    H_iC\widehat\xi_{\mathrm{e}}(v)+d_i^{\mathrm f}\mu_{\mathrm f}(v)&\leq h_i,\\
    H_iC\widehat\xi_{\mathrm{e}}(v)+d_i^{\mathrm f}\tau_{N_{\mathrm{p}}}^{\mathrm f}&\leq h_i.
\end{aligned}
\]
These continuous inequalities also show that the full-history
sets are closed.
Write $\mathcal S_{N_{\mathrm{p}}}^{\mathrm o}(R)\coloneqq\mathcal S_{N_{\mathrm{p}}}(R)$
and $\mathcal S_{N_{\mathrm{p}}}^{\mathrm o}\coloneqq\mathcal S_{N_{\mathrm{p}}}$ for the
output-history construction. Then
\begin{equation}
    \mathcal S_{N_{\mathrm{p}}}^{\mathrm f}(R)
        \subseteq\mathcal S_{N_{\mathrm{p}}}^{\mathrm o}(R)
        \quad\text{for every }R\geq0,
    \qquad
    \mathcal S_{N_{\mathrm{p}}}^{\mathrm f}
        \subseteq\mathcal S_{N_{\mathrm{p}}}^{\mathrm o}.
    \label{eq:matched-admissibility-inclusion}
\end{equation}

To compare the propagation bounds, restrict the operators to the
zero-reference-error subspace, which is invariant under every
$\mathsf{F}_M$. This gives
$a\leq a_{\mathrm f}$ and $\delta_A\leq\delta_{\mathrm f}$,
so $q\leq q_{\mathrm f}$.
The local mismatch and disturbance vectors lie in the same
subspace. Hence
$\beta=\beta_{\mathrm f}$, $\omega=\omega_{\mathrm f}$,
and $\eta=\eta_{\mathrm f}$ at matching arguments, and radius
induction gives $\rho_j\leq\rho_j^{\mathrm f}$.

For the row and terminal tests, restriction and nonexpansive projection in
\eqref{eq:matched-history-norms} give $d_i=d_i^{\mathrm f}$.
Moreover, $u_{N_{\mathrm{p}}}=u_{\mathrm{e}}(v)$, so the nominal terminal
displacement has zero reference component and
$\tau_{N_{\mathrm{p}}}\leq\tau_{N_{\mathrm{p}}}^{\mathrm f}$.
Also $\mu(v)\leq\mu_{\mathrm f}(v)$.
Every output-history test therefore follows from its matched
full-history test. Taking the union over $R$ proves the
second inclusion.

For either matched fixed-radius pair or the two
radius-eliminated sets, consider the same history, previous
reference, and request. If the full-history hold pair is
feasible, its feasible reference fractions form a subset of
those for the output-history construction. Both fraction sets
are nonempty closed subsets of $[0,1]$, by closedness of the
admissible sets and continuity of the reference segment.
Their attained maxima therefore satisfy
$\alpha_{\mathrm o}^{\max}\geq\alpha_{\mathrm f}^{\max}$.
The comparison is pointwise. Both certificates use the same
history, previous reference, request, and fixed design choices.

\subsection{Robust reference governor}

Under the actual history dynamics, a certified applied reference
remains admissible at the next measured history by held-reference
invariance. The DDRG can therefore use that reference as a feasible
hold at the next update. To move toward the current request, the
DDRG tests candidates along the segment from the previous applied
reference to the request. A fraction of zero holds the previous
reference, and a fraction of one applies the request in full.

\begin{theorem}[Recursive feasibility and constraint satisfaction]
\label{thm:rg}
Suppose the problem assumptions and the conditions of
Proposition~\ref{prop:admissibility} hold.
Use its fixed admissible set $\mathcal S_{N_{\mathrm{p}}}$ from
\eqref{eq:finite-horizon-admissibility}.
To make the initial hold feasible, require the pair of the actual
measured history $\xi_0$ and previous applied reference $v_{-1}$
to satisfy
\begin{equation}
    (\xi_0,v_{-1})\in\mathcal S_{N_{\mathrm{p}}}.
    \label{eq:rg-initialization}
\end{equation}
For any requested-reference sequence $r_k\in\mathbb R^m$,
the admissibility test pairs each candidate with the current measured
history $\xi_k$. The fractions that pass form the set
\begin{equation}
    \mathcal A_k
    \coloneqq
    \left\{
        \alpha\in[0,1]:
        \bigl(\xi_k,
        v_{k-1}+\alpha(r_k-v_{k-1})\bigr)
        \in\mathcal S_{N_{\mathrm{p}}}
    \right\},
    \label{eq:rg-feasible-updates}
\end{equation}
Choose and apply the largest certified update,
\begin{equation}
    \alpha_k=\max\mathcal A_k,
    \qquad
    v_k=v_{k-1}+\alpha_k(r_k-v_{k-1}).
    \label{eq:rg-update}
\end{equation}

For every fixed true model $\Theta\in\mathcal P$ and every
disturbance sequence with $w_k\in\mathcal W$, the following
properties hold for all $k\geq0$:
\begin{enumerate}
    \item $\mathcal A_k$ is nonempty and compact, and its
    maximum is attained.

    \item The DDRG is recursively feasible, with
    \[
        (\xi_k,v_{k-1})\in\mathcal S_{N_{\mathrm{p}}},
        \qquad
        (\xi_k,v_k)\in\mathcal S_{N_{\mathrm{p}}},
    \]
    and the physical output and applied reference satisfy
    \begin{equation}
        Hy_k\leq h,
        \qquad v_k\in\mathcal V.
        \label{eq:rg-safety}
    \end{equation}

    \item The update selects the largest certified fraction
    along the reference segment and minimizes the distance
    to the request among its feasible points, with
    \[
        \|v_k-r_k\|_2
        =(1-\alpha_k)\|v_{k-1}-r_k\|_2.
    \]
    If $(\xi_k,r_k)\in\mathcal S_{N_{\mathrm{p}}}$, then
    $\alpha_k=1$ and $v_k=r_k$.
\end{enumerate}
Recursive feasibility and constraint satisfaction also hold
for any selection $\alpha_k\in\mathcal A_k$, including the
hold-reference choice $\alpha_k=0$.
If $\mathcal V$ is convex, then
\[
    \mathcal A_k=[0,\max\mathcal A_k].
\]
\end{theorem}

\begin{proof}
\textit{Hold feasibility and attainment.}
Suppose $(\xi_k,v_{k-1})\in\mathcal S_{N_{\mathrm{p}}}$.
The choice $\alpha=0$ is feasible, so $\mathcal A_k$ is
nonempty. The segment parametrization
\[
    \alpha\longmapsto
    \bigl(\xi_k,v_{k-1}+\alpha(r_k-v_{k-1})\bigr)
\]
is continuous. By Proposition~\ref{prop:admissibility},
$\mathcal S_{N_{\mathrm{p}}}$ is closed. Its inverse image intersected with
$[0,1]$ is closed and compact. The continuous objective
$\alpha$ consequently attains its maximum on $\mathcal A_k$.

\textit{Propagation of feasibility.}
Every selected feasible fraction gives
$(\xi_k,v_k)\in\mathcal S_{N_{\mathrm{p}}}$.
Proposition~\ref{prop:data} establishes $\Theta\in\mathcal P$,
and the actual successor satisfies
\[
    \xi_{k+1}=\mathsf{F}_\Theta\xi_k+\mathsf{G}_\Theta v_k+Ew_k.
\]
The held-reference invariance in
\eqref{eq:admissible-set-invariance} yields
$(\xi_{k+1},v_k)\in\mathcal S_{N_{\mathrm{p}}}$.
The selected reference becomes the previous applied reference
at the next sample. Its admissibility makes the hold fraction
feasible there. The actual initialized pair in
\eqref{eq:rg-initialization} supplies the base case. Induction
then gives existence of the update and recursive feasibility
at every sample.

\textit{Sampled constraints.}
Membership in $\mathcal S_{N_{\mathrm{p}}}$ enforces $v_k\in\mathcal V$.
Its step-zero prediction has $\widehat z_0=z_k$ and
$\rho_0=0$, so its stage test gives
\[
    H_iC_{\mathrm{o}}z_k=H_iy_k\leq h_i.
\]
This proves \eqref{eq:rg-safety}. The argument uses feasibility
of the chosen fraction, so it also proves the assertion for
any $\alpha_k\in\mathcal A_k$.

\textit{Distance to the request.}
For every $\alpha\in[0,1]$,
\[
    \|v_{k-1}+\alpha(r_k-v_{k-1})-r_k\|_2
    =(1-\alpha)\|v_{k-1}-r_k\|_2.
\]
The right-hand side is nonincreasing in $\alpha$, which proves
the distance-minimization assertion. A feasible requested
reference makes $1\in\mathcal A_k$ and is therefore applied
in full.

\textit{Convex reference set.}
If $\mathcal V$ is convex,
Proposition~\ref{prop:admissibility} makes
$\mathcal S_{N_{\mathrm{p}}}$ convex. Its inverse image under the affine
reference map is convex. A nonempty compact convex subset of
$[0,1]$ containing zero is the interval
$[0,\max\mathcal A_k]$.
\end{proof}

The requested reference may vary arbitrarily. The certificate
constrains the applied reference. When $\mathcal V$ is convex,
an interval search can retain a certified feasible lower
endpoint throughout. The exact maximum gives the largest certified
step on the segment. A verified feasible fraction suffices for
recursive feasibility and sampled constraint satisfaction.
Holding the previous reference is the feasible fallback after
successful initialization.

\section{Bounded output-measurement noise}
\label{sec:measurement-noise}

The physical output obeys \eqref{eq:system}. With online sensor error
$\nu_k\in\mathbb R^p$, the DDRG observes
\begin{equation}
    \widetilde y_k=y_k+\nu_k.
    \label{eq:noise-measurement}
\end{equation}
Applied references remain exact. The supplied deterministic bounds distinguish
the offline errors $\nu_{\mathrm{d},t}\in\mathbb R^p$ from the online
errors. Assume
$\|\nu_{\mathrm{d},t}\|_\infty\le\epsilon_{\mathrm{d}}$ for every offline sample and
$\|\nu_k\|_\infty\le\epsilon_{\mathrm{o}}$ for every online sample, including all
stored samples $k=-L,\ldots,0$ at initialization, where
$\epsilon_{\mathrm{d}},\epsilon_{\mathrm{o}}\ge0$ and $L=\ell-1$.
The same physical coefficient matrix and process bound
$\|w_k\|_\infty\le\bar w$ apply during collection and operation.
When the fixed feedback controller also uses noisy sensor readings,
the physical recurrence and process bound must cover their effect on
the physical dynamics.

The noisy record is used to select the model set for prediction.
The DDRG uses this fixed set to propagate measured-history tubes and
bound future physical outputs. Physical safety at initialization
requires a separate certificate.

\subsection{Noisy data and a consistent nominal model}

To check a model against the noisy record, first express the recurrence
in terms of measured outputs. Substituting $y=\widetilde y-\nu$ into
\eqref{eq:system} gives
\begin{equation}
    \widetilde y_{k+1}
    =\Theta\widetilde\phi_k+w_k+\nu_{k+1}
      -\sum_{j=0}^{L}F_j(\Theta)\nu_{k-j},
    \label{eq:noise-regression}
\end{equation}
where $\widetilde\phi_k\in\mathbb R^d$ replaces the output entries of
$\phi_k$ by their measurements and retains the exact reference
entries. The next sample contributes its error directly to the target.
Errors in the output history enter through their lag coefficients.

For a candidate coefficient matrix $M$, the block $\mathscr A(M)$
collects these output-lag coefficients in the order of
Proposition~\ref{prop:data}. The absolute row sum $a_i(M)$ bounds the
amplification of history errors in the corresponding output component.
\[
\begin{aligned}
    \mathscr A(M)&\coloneqq[F_0(M)\ \cdots\ F_L(M)]
    \in\mathbb R^{p\times\ell p},&
    a_i(M)&\coloneqq\|\operatorname{row}_i\mathscr A(M)\|_1.
\end{aligned}
\]
Stack the measured targets and regressors from the $N_{\mathrm{d}}$
valid offline transitions into $\widetilde Y\in\mathbb R^{p\times N_{\mathrm{d}}}$
and $\widetilde\Phi\in\mathbb R^{d\times N_{\mathrm{d}}}$ in the
transition order of Proposition~\ref{prop:data}. The exact consistency
equation for a candidate $M\in\mathcal K$ is
\begin{equation}
    \widetilde Y-M\widetilde\Phi
    =W+N_+(\nu_{\mathrm{d}})-\mathscr A(M)N_-(\nu_{\mathrm{d}}),
    \label{eq:noise-data-equation}
\end{equation}
where $W\in\mathbb R^{p\times N_{\mathrm{d}}}$ contains the process
disturbance witnesses. The matrices
$N_+\in\mathbb R^{p\times N_{\mathrm{d}}}$ and
$N_-\in\mathbb R^{\ell p\times N_{\mathrm{d}}}$ collect target and
output-regressor errors from one sequence $\nu_{\mathrm{d},0:N}$ in
their original time order.
\[
\begin{aligned}
    N_+&\coloneqq[\,\nu_{\mathrm{d},\ell}\ \cdots\ \nu_{\mathrm{d},N}\,],\\
    N_-&\coloneqq[\,\operatorname{col}(\nu_{\mathrm{d},k},\ldots,\nu_{\mathrm{d},k-L})\,]
            _{k=\ell-1}^{N-1}.
\end{aligned}
\]
Each occurrence of a sample uses the same error, whether that sample
appears in the target or in a regressor. With all supplied coefficient
restrictions included in the closed prior set $\mathcal K$, define
$\mathcal P_{\mathrm{EIV}}$ as the set of $M\in\mathcal K$ for
which \eqref{eq:noise-data-equation} has witnesses satisfying
$\|\nu_{\mathrm{d},t}\|_\infty\le\epsilon_{\mathrm{d}}$ for every offline
sample and $\|W\|_{\max}\le\bar w$.
The actual sensor errors and disturbances give feasible witnesses, so
$\Theta\in\mathcal P_{\mathrm{EIV}}$.
To see that the set is closed, take a convergent sequence of compatible
matrices. The finite noise and disturbance boxes supply a convergent
witness subsequence. Continuity preserves the consistency equation at
the limit. Convexity and boundedness require separate justification.

Bounding each residual entry separately yields the outer set
$\mathcal P_{\mathrm{res}}$ of matrices $M\in\mathcal K$ satisfying
\begin{equation}
    |(\widetilde Y-M\widetilde\Phi)_{it}|
    \le\bar w+\epsilon_{\mathrm{d}}[1+a_i(M)]
    \qquad\forall i,t.
    \label{eq:noise-residual-outer}
\end{equation}
The triangle inequality gives
$\mathcal P_{\mathrm{EIV}}\subseteq\mathcal P_{\mathrm{res}}$.
The relaxation drops the shared-error constraints that couple transitions.

Under polyhedral restrictions, either of the following constructions
allows a boundedness check. If $\mathcal K$ includes
valid prior bounds $a_i(M)\le\Lambda_i$, impose fixed residual bounds
$b_{\mathrm{d},i}=\bar w+\epsilon_{\mathrm{d}}(1+\Lambda_i)$.
For polyhedral restrictions this gives a polyhedron. Its boundedness can
be checked by coordinate minimum and maximum linear programs. Bounds
on output coefficients alone need not bound reference coefficients.
Alternatively, for polyhedral $\mathcal K$, fix the signs of the
$\ell p^2$ output-coefficient entries in
\eqref{eq:noise-residual-outer}. Each resulting piece is polyhedral
because $a_i(M)$ is linear within that orthant. Complete feasibility and
boundedness checks over at most $2^{\ell p^2}$ pieces certify a compact
union when every nonempty piece is bounded. The union or its convex hull
can then be used. Known signs and structural parameterizations reduce
this calculation. An unbounded relaxation does not establish that the
exact consistency set is unbounded.

For propagation, fix a certified nonempty compact model set
$\mathcal P_\nu$ that retains all exactly consistent models, so that
\begin{equation}
    \mathcal P_{\mathrm{EIV}}\subseteq\mathcal P_\nu,
    \qquad \Theta\in\mathcal P_\nu.
    \label{eq:noise-model-containment}
\end{equation}
Compactness needs independent certification because measurement errors
can create apparent excitation even when $\widetilde\Phi$ has full row
rank. Structural restrictions can establish boundedness with
rank-deficient data.

Select and fix $\widehat\Theta_\nu\in\mathcal P_{\mathrm{EIV}}$, retaining
a verified shared-error witness for \eqref{eq:noise-data-equation}.
For a proposed coefficient matrix, verify the witness by solving a
linear feasibility problem in the errors and disturbances. Selecting
the coefficients and the witness jointly is generally bilinear.
Apply the witness check to any outer-set fit before using it as the
exactly consistent nominal. The shared-error witness certifies
membership in $\mathcal P_{\mathrm{EIV}}$, which also gives the
membership in $\mathcal P_\nu$ used by the safety argument.

\subsection{Measured-history dynamics and sensor completion}

To predict from the available measurements, use the measured-output
history $\widetilde z_k\in\mathbb R^{n_{\mathrm{o}}}$ and the exact
applied-reference history $u_k$ in the surrogate history
\[
    \widetilde z_k
    \coloneqq\operatorname{col}(\widetilde y_k,\ldots,\widetilde y_{k-L}),
    \qquad
    \widetilde\xi_k\coloneqq\operatorname{col}(\widetilde z_k,u_k)
    \in\mathbb R^n.
\]
Equation~\eqref{eq:noise-regression} gives the surrogate dynamics with
the operators of Proposition~\ref{prop:tube}. For a candidate
$M\in\mathcal P_\nu$, these dynamics take the form
\[
\begin{aligned}
    \widetilde z^+
    &=\mathsf A_M\widetilde z+\mathsf D_Mu+\mathsf B_Mv+E_{\mathrm{o}}e,\\
    u^+&=Su+Jv,\qquad e\in\mathcal W_\nu,
\end{aligned}
\]
where $e\in\mathbb R^p$ is the effective disturbance and
$\mathcal W_\nu$ is the forcing ball constructed below. Only the
measured-output block receives the disturbance. The true measured
history follows these dynamics with $M=\Theta$ and the effective
disturbance in \eqref{eq:noise-regression}.

To cover all allowable combinations of process disturbance and
sensor errors, let
$\mathcal N_{\mathrm{o}}\coloneqq\{n\in\mathbb R^p:\|n\|_\infty\le\epsilon_{\mathrm{o}}\}$.
The effective disturbance image is
\begin{equation}
    \mathcal E_\nu(\mathcal P_\nu)\coloneqq
    \bigcup_{M\in\mathcal P_\nu}
    \biggl\{w+n^+-\sum_{j=0}^{L}F_j(M)n_j:
    w\in\mathcal W,\quad n^+,n_0,\ldots,n_L\in\mathcal N_{\mathrm{o}}\biggr\},
    \label{eq:noise-disturbance-image}
\end{equation}
where $n^+$ is the next sensor error and $n_0,\ldots,n_L$ are
the output-history errors in lag order. This set in $\mathbb R^p$
is the continuous image of a compact product.
For the fixed output-history norm, choose a finite certified bound on
the norm of the injected disturbance and use it to define the forcing ball,
\begin{equation}
\begin{aligned}
    \Omega_\nu&\ge
    \max_{e\in\mathcal E_\nu(\mathcal P_\nu)}\|E_{\mathrm{o}}e\|_{\mathrm{o}},&
    \mathcal W_\nu
    &\coloneqq\{e\in\mathbb R^p:\|E_{\mathrm{o}}e\|_{\mathrm{o}}\le\Omega_\nu\}.
\end{aligned}
    \label{eq:noise-forcing-bound}
\end{equation}
Since $E_{\mathrm{o}}$ is injective, $\mathcal W_\nu$ is a compact norm ball.
The ball contains the actual effective disturbance in
\eqref{eq:noise-regression}. It also covers every admissible choice
of the next sensor error with the true model, physical disturbance,
and past sensor errors fixed.
\begin{equation}
    w_k-\sum_{j=0}^{L}F_j(\Theta)\nu_{k-j}+n
    \in\mathcal W_\nu
    \qquad\text{for every }n\in\mathcal N_{\mathrm{o}}.
    \label{eq:noise-completion}
\end{equation}
The physical-safety proof uses all these sensor completions while
keeping the physical successor fixed. The deterministic bounds suffice
even if the actual errors are dependent.

One computable choice bounds each disturbance component before taking
the output-history norm. It first obtains certified bounds
$\lambda_i\ge\max_{M\in\mathcal P_\nu}a_i(M)$ and sets
\[
    \bar e_i=\bar w+\epsilon_{\mathrm{o}}(1+\lambda_i),\qquad
    \Omega_{\mathrm{box}}
    =\max_{|e_i|\le\bar e_i}\|E_{\mathrm{o}}e\|_{\mathrm{o}}.
\]
For example, with the fixed nonsingular diagonal weighting
$\operatorname{diag}(c_1,\ldots,c_{n_{\mathrm{o}}})$, $c_j>0$, the
weighted infinity norm $\|z\|_{\mathrm{o}}=\max_j c_j|z_j|$ gives
$\Omega_{\mathrm{box}}=\max_{1\le i\le p}c_i\bar e_i$.
For polyhedral model sets, signed-row linear programs compute the
amplification maxima. Complete box-corner enumeration computes the last
norm maximum. A direct calculation of
\eqref{eq:noise-disturbance-image} can be tighter. For a compact polytope
or finite union of polytopes, complete model vertices and noise-box
corners give an exact maximum: the norm is convex in the model for fixed
errors and in the joint error vector for a fixed model. Successive vertex
maximization proves the claim. Incomplete vertex samples give lower
estimates, so they cannot certify \eqref{eq:noise-forcing-bound}.

\subsection{Measured-history tubes and admissibility}

The tubes bound the difference between surrogate histories and their
nominal predictions. To construct them, use the fixed nominal matrices
$\widehat{\mathsf A}_\nu=\mathsf A_{\widehat\Theta_\nu}$,
$\widehat{\mathsf D}_\nu=\mathsf D_{\widehat\Theta_\nu}$, and
$\widehat{\mathsf B}_\nu=\mathsf B_{\widehat\Theta_\nu}$, and define
the additive gain for radius propagation. The terminal construction
requires the common contraction condition $q_\nu<1$,
\begin{equation}
\begin{aligned}
    a_\nu&\coloneqq\|\widehat{\mathsf A}_\nu\|_{\mathrm{o}},&
    \delta_{A,\nu}
      &\coloneqq\max_{M\in\mathcal P_\nu}
        \|\mathsf A_M-\widehat{\mathsf A}_\nu\|_{\mathrm{o}},\\
    q_\nu&\coloneqq a_\nu+\delta_{A,\nu}<1.
\end{aligned}
    \label{eq:noise-gain}
\end{equation}
The local mismatch bounds the effect of using the nominal model at a
given history $\xi=\operatorname{col}(z,u)$ and applied reference $v$.
It includes uncertainty in all three coefficient blocks.
\begin{equation}
\begin{split}
    \beta_\nu(\xi,v)
      \coloneqq\max_{M\in\mathcal P_\nu}
      \bigl\|& (\mathsf A_M-\widehat{\mathsf A}_\nu)z\\
       &+(\mathsf D_M-\widehat{\mathsf D}_\nu)u
       +(\mathsf B_M-\widehat{\mathsf B}_\nu)v\bigr\|_{\mathrm{o}}.
\end{split}
    \label{eq:noise-local-bound}
\end{equation}
Continuity in the candidate model and compactness of the model set make
these maxima finite and attained. The local mismatch function is
continuous and convex in $(\xi,v)$, and, at equal $u,v$,
\[
    |\beta_\nu(\operatorname{col}(z,u),v)
      -\beta_\nu(\operatorname{col}(z',u),v)|
    \le\delta_{A,\nu}\|z-z'\|_{\mathrm{o}}.
\]

For a held candidate $v$ and starting history
$\widetilde\xi=\operatorname{col}(\widetilde z,u)$, initialize the nominal
prediction and its radius at
$\widehat z_0=\widetilde z$, $u_0=u$, and $\rho_0^\nu=0$. The stored
measured history is known exactly, which gives zero initial surrogate
radius. Sensor uncertainty remains in the physical output history.
Propagate the reference history, nominal measured-output history, and
radius by
\begin{align}
    u_{j+1}&=Su_j+Jv,\nonumber\\
    \widehat z_{j+1}
      &=\widehat{\mathsf A}_\nu\widehat z_j
        +\widehat{\mathsf D}_\nu u_j+\widehat{\mathsf B}_\nu v,
        \nonumber\\
    \rho_{j+1}^\nu
      &=q_\nu\rho_j^\nu
        +\beta_\nu(\operatorname{col}(\widehat z_j,u_j),v)
        +\Omega_\nu.
    \label{eq:noise-tube-recursion}
\end{align}
The tube varies only the measured-output block,
\[
    \Gamma_j^\nu(\widetilde\xi,v)
    \coloneqq\{\operatorname{col}(\widehat z_j+e,u_j):
         \|e\|_{\mathrm{o}}\le\rho_j^\nu\}\subseteq\mathbb R^n,
\]
where the deviation $e$ in this definition lies in
$\mathbb R^{n_{\mathrm{o}}}$. For constraint row $i$, the coefficient
$d_i$ converts the history radius to row support, as in
Proposition~\ref{prop:margins}, and $s_i$ is the support of the
online sensor-error ball in the same constraint direction.
\[
    d_i=\|H_iC_{\mathrm{o}}\|_{\mathrm{o},*},\qquad
    s_i\coloneqq\epsilon_{\mathrm{o}}\|H_i\|_1.
\]

To certify continuation, center the terminal region at the nominal
equilibrium $\widehat z_{\mathrm{e},\nu}(v)$ with reference history
$u_{\mathrm{e}}(v)$. The mismatch $\eta_\nu(v)$ at this equilibrium
and the forcing bound determine the radius floor $\mu_\nu(v)$.
The bound $\tau_{\nu,j}(\widetilde\xi,v)$ adds the tube radius to
the nominal center's distance from the equilibrium.
\begin{align}
    \widehat z_{\mathrm{e},\nu}(v)
      &\coloneqq(I-\widehat{\mathsf A}_\nu)^{-1}
        (\widehat{\mathsf D}_\nu u_{\mathrm{e}}(v)
         +\widehat{\mathsf B}_\nu v),\nonumber\\
    \eta_\nu(v)
      &\coloneqq\beta_\nu(\operatorname{col}
         (\widehat z_{\mathrm{e},\nu}(v),u_{\mathrm{e}}(v)),v),\nonumber\\
    \mu_\nu(v)&\coloneqq\frac{\eta_\nu(v)+\Omega_\nu}{1-q_\nu},
        \nonumber\\
    \tau_{\nu,j}(\widetilde\xi,v)
      &\coloneqq\|\widehat z_j-\widehat z_{\mathrm{e},\nu}(v)\|_{\mathrm{o}}
        +\rho_j^\nu,\qquad j\ge0.
    \label{eq:noise-terminal-quantities}
\end{align}
The inverse exists because $a_\nu\le q_\nu<1$.

For an integer $N_{\mathrm{p}}\ge\max\{1,L\}$, the surrogate
admissibility tests use $h_i+s_i$ as the limit in row $i$.
The physical safety proof recovers $h_i$ by maximizing over sensor
completions. Let $\mathcal S^\nu_{N_{\mathrm{p}}}\subseteq
\mathbb R^n\times\mathbb R^m$ consist of pairs $(\widetilde\xi,v)$ with
$v\in\mathcal V$ satisfying, for every constraint row,
\begin{align}
    H_iC_{\mathrm{o}}\widehat z_j+d_i\rho_j^\nu
      &\le h_i+s_i,&&0\le j<N_{\mathrm{p}},
        \label{eq:noise-stage-tests}\\
    H_iC_{\mathrm{o}}\widehat z_{\mathrm{e},\nu}(v)+d_i\mu_\nu(v)
      &\le h_i+s_i,
        \label{eq:noise-terminal-floor}\\
    H_iC_{\mathrm{o}}\widehat z_{\mathrm{e},\nu}(v)
       +d_i\tau_{\nu,N_{\mathrm{p}}}(\widetilde\xi,v)
      &\le h_i+s_i.
        \label{eq:noise-terminal-entry}
\end{align}
Write $s\in\mathbb R^{n_c}$ for the vector of rowwise support shifts.

\subsection{Physical safety and recursive feasibility}

At each update the DDRG tests references using the measured history
and the surrogate limits. The initial physical-output check supplies
safety at the first sample. Subsequent physical safety follows from the
preceding accepted update by considering all sensor completions.

\begin{theorem}[Reference governor with bounded measurement noise]
\label{thm:rg-noise}
Assume the physical and error bounds above, a fixed nonempty compact
$\mathcal P_\nu$ satisfying \eqref{eq:noise-model-containment}, the fixed
consistent nominal and norm, the forcing bound
\eqref{eq:noise-forcing-bound}, and \eqref{eq:noise-gain}.
Let $\mathcal V$ be compact and $N_{\mathrm{p}}\ge\max\{1,L\}$.
Initialize with
$(\widetilde\xi_0,v_{-1})\in\mathcal S^\nu_{N_{\mathrm{p}}}$ and certified
physical initial safety $Hy_0\le h$.
A sufficient measurement-based initialization test is
\begin{equation}
    H_i\widetilde y_0+s_i\le h_i\qquad\forall i.
    \label{eq:noise-initial-safety}
\end{equation}
This physical-safety check is imposed once; subsequent updates use
membership in $\mathcal S^\nu_{N_{\mathrm{p}}}$.
For each requested reference $r_k$, choose the largest certified
fraction along the segment from the previously applied reference to
the request,
\begin{equation}
\begin{split}
    \alpha_k
    &=\max\{\alpha\in[0,1]: (\widetilde\xi_k,
        v_{k-1}+\alpha(r_k-v_{k-1}))\in\mathcal S^\nu_{N_{\mathrm{p}}}\},\\
    v_k&=v_{k-1}+\alpha_k(r_k-v_{k-1}).
\end{split}
    \label{eq:noise-governor}
\end{equation}
Then the following conclusions hold.
\begin{enumerate}
    \item The maximum exists at every step, and the measured pair
          remains recursively feasible.
    \item The constraints $Hy_k\le h$ and $v_k\in\mathcal V$ hold
          for every $k\ge0$. Every certified feasible fraction has
          the same safety guarantee.
    \item If $\mathcal V$ is convex, the feasible fractions form an
          interval containing zero.
\end{enumerate}
\end{theorem}

\begin{proof}
\textit{Containment and reset.}
Equation~\eqref{eq:noise-model-containment} supplies true-model
membership. The model set, nominal, norm, and forcing bound are fixed,
and the reference history is exact. The bound
$\|\mathsf A_M\|_{\mathrm{o}}\le q_\nu$ and the output-history Lipschitz
constant $\delta_{A,\nu}$ for the local bound give the two gain
estimates used in Proposition~\ref{prop:tube}. That proposition bounds
the disturbance through its injection norm, so
\eqref{eq:noise-forcing-bound} supplies the required estimate on the
compact set $\mathcal W_\nu$.
In particular, trajectories lie in $\Gamma_j^\nu$, and the entire tubes
restarted from any admissible successor satisfy
\[
    \Gamma_j^\nu(\widetilde\xi^+,v)
    \subseteq\Gamma_{j+1}^\nu(\widetilde\xi,v).
\]
If $b_j$ bounds the center displacement at step $j$, the additive gain
accounts for $a_\nu b_j$ in the next center displacement and
$\delta_{A,\nu}b_j$ in the restarted radius.
The norm-ball support calculation in Proposition~\ref{prop:margins}
then gives the surrogate envelope
$H_iC_{\mathrm{o}}\widehat z_j+d_i\rho_j^\nu$.

\textit{Terminal continuation and hold invariance.}
For $j\ge L$, $u_j=u_{\mathrm{e}}(v)$, and the terminal-distance bound obeys
\[
    \tau_{\nu,j+1}
    \le q_\nu\tau_{\nu,j}+\eta_\nu(v)+\Omega_\nu.
\]
The terminal tests supply the nonnegative radius
$R_* =\max\{\mu_\nu(v),\tau_{\nu,N_{\mathrm{p}}}\}$.
The terminal fiber of this radius contains the complete terminal tube
and remains invariant under the held reference. Reset inclusion
preserves the earlier stage tests, and terminal continuation supplies the
new end of the prediction. Together they prove hold-reference
invariance of $\mathcal S^\nu_{N_{\mathrm{p}}}$.
Continuity of the prediction quantities makes this set closed. When
$\mathcal V$ is convex, their convexity makes the set convex; convexity
of $\mathcal P_\nu$ is unnecessary.

\textit{Physical safety from sensor completion.}
Fix an accepted pair $(\widetilde\xi_k,v_k)$ and the actual physical
successor $y_{k+1}$. For every $n\in\mathcal N_{\mathrm{o}}$,
\eqref{eq:noise-completion} gives a surrogate successor whose output
history is
\[
    \operatorname{col}
    (y_{k+1}+n,\widetilde y_k,\ldots,\widetilde y_{k-L+1}),
\]
with the exact updated reference history; when $L=0$ only the first
output block remains. Hold invariance places each such successor in
$\mathcal S^\nu_{N_{\mathrm{p}}}$. Its stage-zero constraint gives
\[
    H_i y_{k+1}+H_i n\le h_i+s_i
    \qquad\forall n\in\mathcal N_{\mathrm{o}}.
\]
Take the supremum over $n$ while keeping the physical successor fixed.
In each row, the maximum sensor contribution equals that row's
support shift. It cancels the shift on the right and yields
$H_i y_{k+1}\le h_i$. The maximizer may differ between rows because
each row inequality holds for every $n$. Choosing the actual
$n=\nu_{k+1}$ also proves
$(\widetilde\xi_{k+1},v_k)\in\mathcal S^\nu_{N_{\mathrm{p}}}$.

\textit{Feasible updates and initialization.}
The accepted pair therefore makes the next hold fraction feasible.
The assumed initial pair makes holding feasible at the first update.
The feasible fractions in
\eqref{eq:noise-governor} are a nonempty closed subset of $[0,1]$, so
their maximum is attained. The argument of Theorem~\ref{thm:rg} then
establishes the induction. Under convexity of $\mathcal V$, the same
affine-segment argument gives an interval of feasible fractions
containing zero. The separate initial-safety assumption covers $k=0$.
The test \eqref{eq:noise-initial-safety} is sufficient because
$y_0=\widetilde y_0-\nu_0$. Subsequent physical safety follows from the
preceding accepted prediction, so the initialization test need not
remain satisfied.
\end{proof}

\subsection{Physical margins and fixed-design comparisons}

The sensor-completion argument also converts the surrogate tube
envelope into a bound on each future physical output. Fix an admissible
physical rollout under a held reference and a step $j\ge1$. Keep all
sensor errors before its final sample unchanged
and let the final error range over $\mathcal N_{\mathrm{o}}$.
Every resulting measured sequence satisfies the surrogate recurrence
with forcing in $\mathcal W_\nu$. Tube containment therefore gives
$y_{k+j}+\mathcal N_{\mathrm{o}}\subseteq C\Gamma_j^\nu$.
The norm-ball support calculation and maximization over the final
sensor value yield
\begin{equation}
    H_i y_{k+j}
    \le H_iC_{\mathrm{o}}\widehat z_j+d_i\rho_j^\nu-s_i,
    \qquad j\ge1.
    \label{eq:noise-physical-envelope}
\end{equation}
The future physical margin is $d_i\rho_j^\nu-s_i$. The initial output
is certified separately, with \eqref{eq:noise-initial-safety} giving a
sufficient measurement-based test. Early surrogate histories can
contain old measured samples, so \eqref{eq:noise-physical-envelope}
applies to their current-output projection. The margin is nonnegative
because
$\mathcal N_{\mathrm{o}}\subseteq\mathcal E_\nu(\mathcal P_\nu)$ gives
$s_i\le d_i\Omega_\nu$, and $\rho_j^\nu\ge\Omega_\nu$ for $j\ge1$.

At fixed model set, nominal, norm, forcing bound, and horizon, replacing
the surrogate right-hand side successively by $h-s$, $h$, and $h+s$
gives nested admissible sets in that order. With the same physical
initialization requirement, the construction above therefore admits
every reference update admitted by either smaller set.
Likewise, the exact direct-image maximum is no greater than
$\Omega_{\mathrm{box}}$. Reducing this certified constant decreases
every positive-step tube radius and the terminal floor. The initial
radius remains zero. The reduction enlarges the admissible set
without changing the online recursion.

The offline bound affects which coefficients the model set retains and
how much mismatch it permits. The online bound affects the effective
disturbance, the support shift, and the measured initial history.
The retained coefficients and propagation gain couple their effects
on the physical margin.

When the supplied bounds satisfy $\epsilon_{\mathrm{d}}=\epsilon_{\mathrm{o}}=0$, the noisy
data identity becomes the exact-measurement identity,
$\mathcal E_\nu(\mathcal P_\nu)=\mathcal W$, and $s=0$.
With matching model-set, nominal, norm, and disturbance-bound choices,
the certificate and DDRG reduce to the preceding exact-measurement
construction.

\clearpage
\section{Numerical certification}
\label{num:section}

On one benchmark, we compare how much prediction disagreement remains
as the record grows, how much constraint margin the prediction tube
uses, and how far the DDRG can move toward a request in a pointwise
comparison and in closed-loop runs.

\subsection{Numerical benchmark}
\label{num:setup}

The benchmark is the coupled lag-two family
\begin{equation}
 y_{k+1}=\vartheta K y_{k-1}
              +(I_2-\vartheta K)v_{k-1}+w_k,
 \label{num:plant}
\end{equation}
where the output and reference each have two components,
$K\in\mathbb R^{2\times2}$ is known, and $\vartheta$ is an unknown scalar.
In the coefficient order
of \eqref{eq:regression}, the parameterization is
\begin{equation}
 \Theta(\vartheta)
 =\begin{bmatrix}0_{2\times2}&\vartheta K&0_{2\times2}&I_2-\vartheta K\end{bmatrix}.
 \label{num:parameterization}
\end{equation}
The prior is the image of
$[\underline\vartheta,\overline\vartheta]$ under this parameterization, with
\[
 K=\begin{bmatrix}\frac12&\frac14\\[2pt]-\frac14&\frac38\end{bmatrix},
 \qquad
 (\underline\vartheta,\overline\vartheta,\vartheta_\star)
 =\left(\frac18,\frac34,\frac13\right).
\]
The induced-norm bound $\overline\vartheta\|K\|_\infty<1$ ensures
stability of the homogeneous two-step recurrence. The simulator and
true-model evaluator alone use $\vartheta_\star$. All models have
$G_{\mathrm{dc}}=I_2$, common equilibrium histories, and $\eta(v)=0$.
We set $T=I_2$ and compare transient disagreement and propagation.

The benchmark uses the physical constraints in \eqref{eq:constraints}
and the disturbance bound in \eqref{eq:disturbance}, with the fixed
limits and bound given in Table~\ref{num:parameters}. The matched norms are
\begin{equation}
 \begin{aligned}
  \|z\|_{\mathrm o}
    &=\|\operatorname{diag}(I_2,\gamma I_2)z\|_\infty,\\
  \|\operatorname{col}(z,u)\|_{\mathrm f}
    &=\max\{\|z\|_{\mathrm o},\kappa\|u\|_\infty\},
 \end{aligned}
 \label{num:norms}
\end{equation}
With the weights in Table~\ref{num:parameters}, these norms satisfy
\eqref{eq:matched-history-norms}. Signals are dimensionless. Time is
measured in samples.

The three prefixes of one deterministic offline record have
$n_s=4,8,12$ valid transitions, respectively, for $s=1,2,3$.
Intersecting the prior
with the scalar residual inequalities gives intervals
$[\underline\vartheta_s,\overline\vartheta_s]$ with matrix images
$\mathcal P_s$. The longest-record midpoint defines the common nominal,
$\widehat\Theta=\Theta(\widehat\vartheta)$ with
$\widehat\vartheta=(\underline\vartheta_3+\overline\vartheta_3)/2$.
This nominal belongs to every $\mathcal P_s$. The common comparison history is
$\xi_{\mathrm{eval}}=0_6$ with previous reference zero. The held probe,
horizon, comparator radius, and sensor bounds are in
Table~\ref{num:parameters}. Appendix~\ref{num:appendix}
specifies the deterministic signals and certification tolerances.

\subsection{Data informativeness}
\label{num:information}

To compare the prediction disagreement across records at a single
regressor, we use the same prior, feasible nominal, and query
\[
 \phi_{\mathrm{probe}}=\operatorname{col}(0_2,0_2,v_{\mathrm{probe}},v_{\mathrm{probe}}).
\]
This is the nominal regressor one step after applying the held probe from
$\xi_{\mathrm{eval}}$. Interval endpoint evaluation gives the exact
one-step disagreement in \eqref{eq:model-uncertainty} because the norm
of the affine mismatch is convex in $\vartheta$.

\begin{figure}[!htbp]
  \centering
  \includegraphics[width=\linewidth]{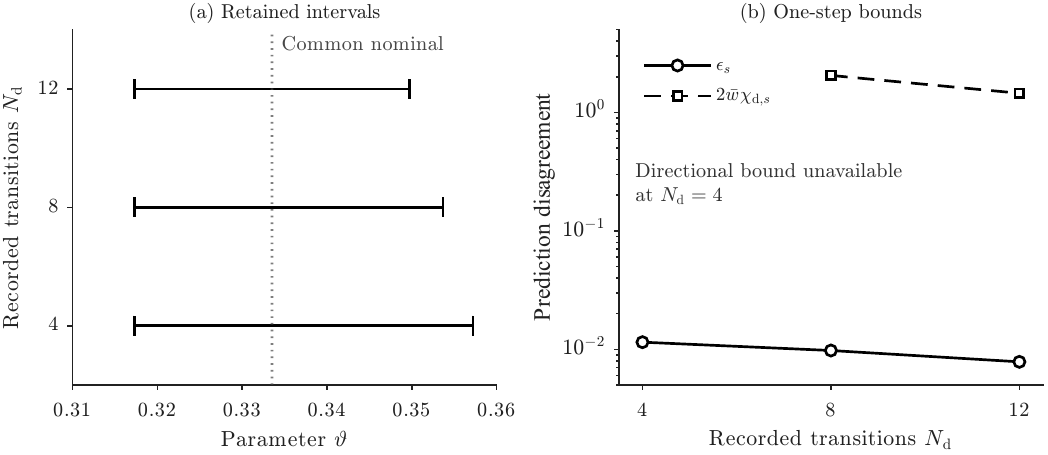}
  \caption{Nested records at a common nominal model and query. (a) Retained
  parameter intervals and the common nominal. (b) One-step disagreement and
  its directional upper bound on a logarithmic scale. At four transitions
  the query lies outside the regressor range, so the bound is unavailable.
  All quantities are dimensionless.}
  \label{num:fig-information}
\end{figure}

Figure~\ref{num:fig-information} compares the three retained parameter
intervals. Rounded outward, they are
\[
 [0.3173,0.3573],\qquad [0.3173,0.3537],\qquad [0.3173,0.3498].
\]
The lower endpoint is the same at all three prefixes.
The one-step disagreement decreases from $0.0115$ to $0.00977$ and
$0.00784$ (rounded). An exact range separator places the four-transition
query outside the recorded regressor range, so the directional bound
from Proposition~\ref{prop:data} is unavailable. The structured disagreement
remains finite on the prior-bounded model set. The two available directional bounds,
approximately $2.05$ and $1.45$, substantially exceed the corresponding
structured disagreements. The numerical rank diagnostics are $4$, $8$,
and $8$ at tolerance $10^{-10}$.

\subsection{Constraint-margin consumption}
\label{num:margins}

To distinguish the contribution of model uncertainty from the contribution
of tube propagation at the same row and step, we evaluated the envelopes of
Proposition~\ref{prop:margins} at the same zero history
$\xi_{\mathrm{eval}}$ and held probe $v_{\mathrm{probe}}$, using the
common nominal model, output-history norm, constraints, and disturbance
bound for all three model sets. The true-model envelope and information loss
use the simulation parameter for evaluation only. For the longest record,
we also evaluated the full-history tube under the matched norm choices
in \eqref{num:norms}.

\begin{figure}[!htbp]
  \centering
  \includegraphics[width=\linewidth]{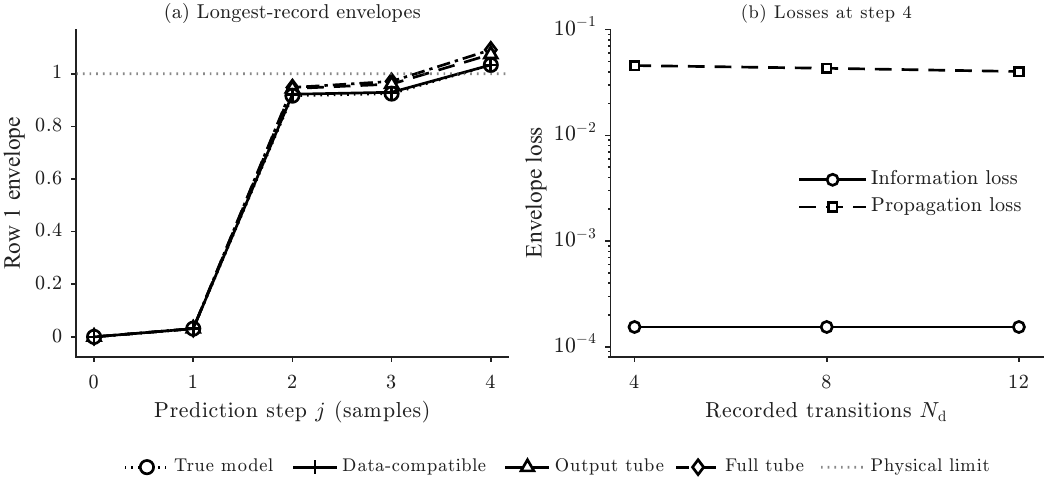}
  \caption{Constraint margins at a common history, held probe, and
  disturbance bound. (a) Row-one envelopes and the physical limit for the
  longest record, before reference modification. (b) Row-one information
  and propagation losses at step four for the three prefixes. The loss axis
  is logarithmic. All quantities are dimensionless.}
  \label{num:fig-margins}
\end{figure}

At row~1 and prediction step~4, the information loss is approximately
$1.54\times10^{-4}$ for every prefix. The propagation loss decreases
from approximately $0.0458$ to $0.0432$ and $0.0402$ as the record
extends from 4 to 8 and 12 transitions. Propagation accounts for most
of the output-history tube's gap to the true-model envelope at this
row and step.
Figure~\ref{num:fig-margins} shows the row-one envelopes for the
longest prefix under the full probe before reference modification.
Across all evaluated rows and steps of that prefix, the full-history
envelope exceeds the output-history envelope by up to approximately
$0.0173$.

The trajectory-containment and whole-tube reset checks passed on
the 20 prescribed fixed-model rollouts. Proposition~\ref{prop:tube}
gives the guarantee for all eligible successors. Appendix~\ref{num:appendix}
records the envelope computation, the rollout suite, and the signed
residuals.

\subsection{Certified reference authority}
\label{num:updates}

Table~\ref{num:authority} compares how far four certificates allow a
reference change toward the same held probe from
$(\xi_{\mathrm{eval}},v_{-1})=(0_6,0_2)$ at $N_{\mathrm{p}}=4$.
All four designs use $\mathcal P_3$ and the common nominal. The norm weights
and comparator radius are fixed across horizons. All four initial holds
were certified.

\begin{table}[!htbp]
\centering
\caption{Pointwise reference authority for the common initial history,
probe, and horizon of four samples. Gains are rounded, and the certified
fraction brackets are rounded outward.}
\label{num:authority}
\small
\setlength{\tabcolsep}{3pt}
\begin{tabular}{@{}>{\raggedright\arraybackslash}p{0.43\linewidth}>{\raggedright\arraybackslash}p{0.14\linewidth}>{\raggedright\arraybackslash}p{0.34\linewidth}@{}}
\toprule
Tube / terminal design & Gain & Certified fraction bracket \\
\midrule
Full history / fixed radius & $0.747$ & $[0.7998,\,0.8008]$ \\
Output history / fixed radius & $0.620$ & $[0.7998,\,0.8008]$ \\
Full history / radius eliminated & $0.747$ & $[0.8251,\,0.8262]$ \\
Output history / radius eliminated & $0.620$ & $[0.8398,\,0.8409]$ \\
\bottomrule
\end{tabular}
\end{table}

The search certified a feasible lower endpoint and an infeasible upper
endpoint with width $2^{-10}$ before display rounding. At the rejected
upper endpoints, the fixed-radius designs failed the terminal-output
condition and the radius-eliminated designs failed terminal entry. The
fixed-radius brackets coincide at the reported resolution. The brackets
resolve an increase under radius elimination for both norms and a further
increase with output history within the radius-eliminated designs. This
comparison is pointwise, as in \eqref{eq:matched-admissibility-inclusion}.
For every design, the brackets also resolve an increase from one to four
samples. From four to eight, the fixed-radius brackets coincide at the
reported resolution, while the radius-eliminated brackets resolve
further increases.

\clearpage
\begin{figure}[!htbp]
  \centering
  \includegraphics[width=\linewidth]{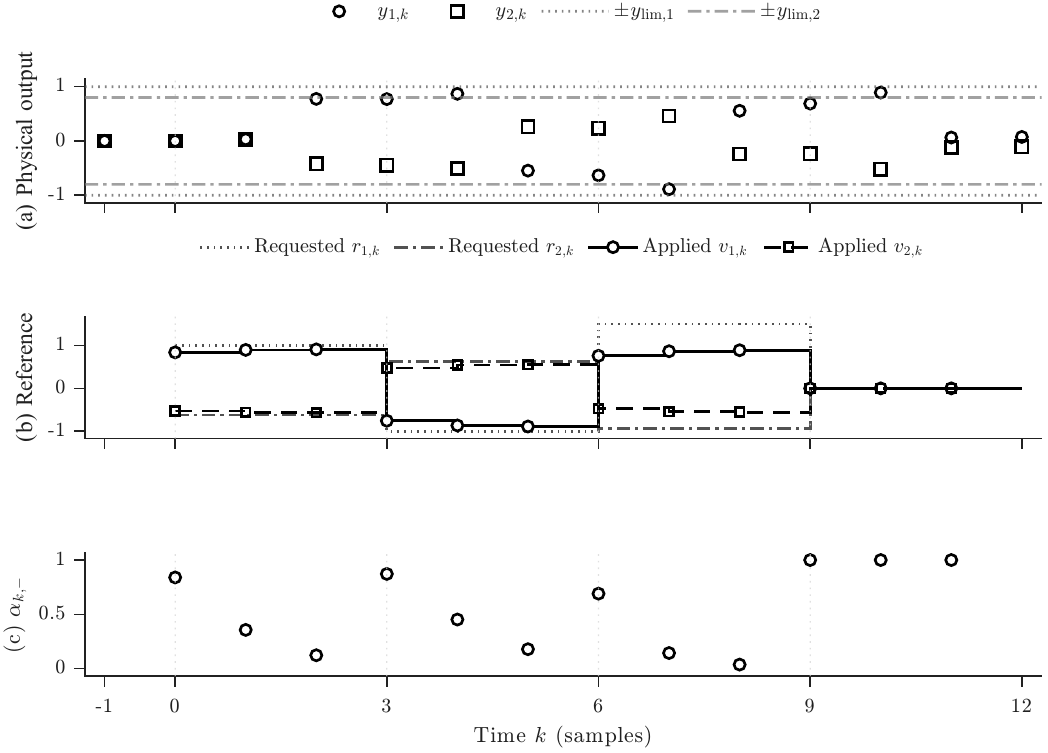}
  \caption{Exact-measurement run of the DDRG with the output-history tube
  and radius elimination over twelve updates. (a) Physical outputs and limits, including
  initialization at $k=-1$ and the final successor at $k=12$. (b) Requested
  and applied references. (c) Applied certified lower fractions
  $\alpha_{k,-}$ from brackets of width at most $2^{-10}$. The fractions
  at $k=10,11$ yield no reference motion because those request segments
  have zero length. Vertical guides mark startup and request changes.
  Time is in samples and signals are dimensionless.}
  \label{num:fig-updates}
\end{figure}

The exact-measurement trace in Figure~\ref{num:fig-updates} uses the DDRG
with the output-history tube and radius elimination. The four requests in
\eqref{num:requests} include one outside $\mathcal V$. The DDRG applied the
certified lower fractions, with a minimum of $0.0371$ (rounded) over nonzero
request segments. The sampled outputs and applied references respected their
limits over all twelve updates, and the next-hold checks passed through the
final successor.

\begin{samepage}
In the bounded-noise run with the sensor bounds in
Table~\ref{num:parameters}, the measured-history hold and separate physical
startup check passed. All twelve updates passed the physical-output,
reference-limit, one-step envelope, and next-hold checks. This run
changes identification, the nominal model, and online sensing jointly
under the same physical plant, limits, disturbance sequence, requests,
and horizon. The noisy nominal satisfied the shared-error consistency
test in \eqref{eq:noise-data-equation}.
\par
\end{samepage}

\section{Conclusion}

With exact measurements and a held reference, output-history tubes
bound each output constraint row for every data-consistent model and
admissible disturbance sequence. The invariant admissible set certifies
indefinite continuation under that reference and preserves a feasible
hold at the next sample. From a certified initial hold, the DDRG
selects the largest certified step along the reference segment while
maintaining recursive feasibility and sampled constraint satisfaction.
This certificate requires fixed choices of a compact data-consistent
model set, a feasible nominal, and an output-history norm with $q<1$.
It also requires an integer horizon $N_{\mathrm{p}}\geq\max\{1,L\}$,
where $L=\ell-1$, so that the held reference fills the applied-reference
history.

Directional prediction disagreement measures one-step model mismatch
at a given regressor, while tube radii bound the accumulated effects of
model mismatch and disturbances. Uncertainty utilization quantifies
the fraction of positive nominal clearance consumed by a tube margin.
Stage, terminal-floor, and terminal-entry slacks identify restrictions
on certified reference changes and support certificate comparisons.
Propagation loss measures the excess of the tube envelope over the
data-compatible envelope. Information loss measures the excess of
that envelope over the true-model envelope and requires the true
plant for evaluation.

The bounded-measurement-noise extension preserves these guarantees using
a compact outer model set containing all models consistent with the
noisy record, a verified consistent nominal, $q_\nu<1$, and a forcing
bound covering process disturbances and sensor errors. Initialization
additionally requires an admissible measured-history hold and separate
certification of physical safety. In the coupled MIMO benchmark,
one-step disagreement decreased across longer nested records at the
fixed probe regressor. Propagation accounted for most of the
tube-to-true-model gap at row~1, step~4, while radius elimination
increased the certified fraction for both norms in the pointwise
comparison. The envelope comparison illustrates the distinction
between retained model uncertainty and propagation conservatism.

\par\vfill\clearpage
\appendix
\section{Numerical parameters and certification}
\label{num:appendix}

\subsection*{Frozen settings and signals}
Table~\ref{num:parameters} records the settings used with the
coupling matrix, parameterization, true parameter, and prior in
Section~\ref{num:setup}.
The output constraints have $H=\operatorname{col}(I_2,-I_2)$ and
$h=\operatorname{col}(y_{\mathrm{lim}},y_{\mathrm{lim}})$.
The reference set is $\mathcal V=\{v:|v_i|\le v_{\mathrm{lim},i},\ i=1,2\}$.

\begin{table}[!htbp]
\centering
\caption{Frozen scientific and numerical settings.}
\label{num:parameters}
\small
\setlength{\tabcolsep}{3pt}
\begin{tabular}{@{}>{\raggedright\arraybackslash}p{.45\linewidth}>{\raggedright\arraybackslash}p{.50\linewidth}@{}}
\toprule
Setting & Value \\
\midrule
Process bound $\bar w$ & $1/32$ \\
Positive output and reference limits & $y_{\mathrm{lim}}=(1,4/5)^\top$, $v_{\mathrm{lim}}=(1,7/8)^\top$ \\
Offline amplitudes $(A_1,A_2)$ and dwells $(P_1,P_2)$ & $(3/4,5/8)$ and $(2,3)$ samples \\
Offline disturbance fraction $c_{\mathrm d}$ & $1/2$ \\
Matched norm weights and common horizon & $(\gamma,\kappa)=(3/5,3)$, $N_\star=4$ samples \\
Held probe and fixed comparator radius & $v_{\mathrm{probe}}=(1,-5/8)^\top$, $R_{\mathrm{fix}}=1/5$ \\
Online request settings & $K_{\mathrm b}=3$ samples, $\lambda_{\mathrm{out}}=3/2$ \\
Offline and online sensor bounds & $(\epsilon_{\mathrm d},\epsilon_{\mathrm o})=(1/256,1/128)$ \\
Fraction and global envelope bracket targets & $\varepsilon_\alpha=2^{-10}$ and $\varepsilon_{\mathrm{glob}}=10^{-6}$ \\
\bottomrule
\end{tabular}
\end{table}

The single offline record starts at $y_{\mathrm d,-1}=y_{\mathrm d,0}=0$
and $v_{\mathrm d,-1}=0$, with
\begin{equation}
 \begin{aligned}
 v_{\mathrm d,k,i}&=A_i(-1)^{\lfloor k/P_i\rfloor},\quad i=1,2,\\
 w_{\mathrm d,k}&=c_{\mathrm d}\bar w\,
       \operatorname{col}((-1)^k,(-1)^{\lfloor k/2\rfloor}),
 \end{aligned}
 \label{num:data-signals}
\end{equation}
For prefix $s=1,2,3$, $N_{\mathrm d}=n_s$ and $N=n_s+1$, where $n_s$ is
the count in Section~\ref{num:setup}. The retained transitions are
$k=1,\ldots,n_s$. The same section specifies the common nominal. Exact
primal and dual witnesses certify the directional bounds for the two
longer prefixes. A range separator certifies that the first query has
no such bound.

\subsection*{Envelope and reset certificates}
Proposition~\ref{prop:margins} defines the compared envelopes and
losses. The fixed-model supports in \eqref{eq:fixed-model-bound}
hold the model fixed across the rollout. Bernstein subdivision
\citep{garloff1993bernstein}, performed in exact rational arithmetic,
yielded matching upper and lower bounds for all $60$ envelope maxima
across three prefixes, four constraint rows, and prediction steps
$0,\ldots,4$.

The disturbance cycle is $w_j=\bar w\,\mathcal C_{1+(j\bmod4)}$, with
$\mathcal C=((1,1)^\top,(1,-1)^\top,(-1,-1)^\top,(-1,1)^\top)$.
The finite suite combined zero disturbance and the four cycle shifts
with $\underline\vartheta_3,\widehat\vartheta,\vartheta_\star$, and
$\overline\vartheta_3$ as fixed model parameters. Each test used the
output-history tube for $\mathcal P_3$ and started from the zero history.
The maximum signed trajectory and whole-tube reset residuals were
both exactly zero across the $20$ tests, with matching reference blocks.

\subsection*{Pointwise and closed-loop checks}
Exact bisection enclosed the maximal feasible fraction. At each update,
the DDRG applied the certified lower endpoint of the bracket.

Both runs used $\vartheta_\star$, $y_{-1}=y_0=v_{-1}=0$, the unshifted
disturbance cycle, and $N_{\mathrm p}=4$. Requests followed
\begin{equation}
 r_k=\begin{cases}
 v_{\mathrm{probe}}, &0\le k<K_{\mathrm{b}},\\
 -v_{\mathrm{probe}}, &K_{\mathrm{b}}\le k<2K_{\mathrm{b}},\\
 \lambda_{\mathrm{out}}v_{\mathrm{probe}}, &2K_{\mathrm{b}}\le k<3K_{\mathrm{b}},\\
 0_2, &3K_{\mathrm{b}}\le k<4K_{\mathrm{b}}.
 \end{cases}
 \label{num:requests}
\end{equation}
For the noise check, the longest offline record used sensor errors
$\nu_{\mathrm{d},t}=\epsilon_{\mathrm d}\mathcal C_{1+(t\bmod4)}$.
The online errors were $\nu_k=\epsilon_{\mathrm o}\mathcal C_{1+(k\bmod4)}$,
including $k=-1,0$. The noisy outer interval intersects the structural
prior with \eqref{eq:noise-residual-outer}, using
$a_i(\Theta(\vartheta))=\vartheta\|\operatorname{row}_iK\|_1$.
Its midpoint had an exact shared-error witness verified by linear
feasibility in \eqref{eq:noise-data-equation}. The forcing certificate
used the upper endpoint $\overline\vartheta_\nu$ in
\begin{equation}
 \Omega_\nu=\max_{i=1,2}
 \left[\bar w+\epsilon_{\mathrm{o}}
  \left(1+\overline\vartheta_\nu
                 \|\operatorname{row}_iK\|_1\right)\right].
 \label{num:noise-forcing}
\end{equation}
The noisy gain and forcing bound were approximately
$(q_\nu,\Omega_\nu)=(0.627,0.0412)$. The measured-history hold with
surrogate limits $h+s$ and the one-time physical startup test
\eqref{eq:noise-initial-safety} passed. The matched zero-noise reduction
passed. Both $12$-update records include the final physical successor.
Each noisy successor passed the one-step physical envelope check.

The maximum signed physical-output and next-hold residuals were
approximately $-0.108$ and $-0.00810$ in the exact-measurement run and
$-0.126$ and $-0.00448$ in the noisy run. On nonzero request segments,
the smallest applied fractions were approximately $0.0371$ and $0.0391$, respectively.
These minima summarize different closed-loop histories.

\section*{Funding Sources}

This research was partially supported by the MEXT SPReAD AI for Science program 2026 (Grant No. 26276994).

\section*{Acknowledgments}

OpenAI ChatGPT was used to assist with manuscript preparation, literature review, and consistency checks. The author retains full responsibility for the scientific content and final manuscript.

\bibliographystyle{plainnat}
\bibliography{references}

\end{document}